\documentclass{aastex63}

\usepackage{CJKutf8}
\usepackage{multirow}%
\usepackage{tabularx}
\usepackage{threeparttable}
\usepackage{amsmath}
\usepackage{appendix}
\usepackage{xcolor}  
\usepackage{tikz}  
\usetikzlibrary{arrows,shapes,chains}

\makeatletter

\newcommand{\Rmnum}[1]{\expandafter\@slowromancap\romannumeral #1@}

\makeatother

\newcommand{\ds}{\color{black}}
\newcommand{\dsn}{\color{black}}
\newcommand{\jw}{\color{black}}
\newcommand{\dsnn}{\color{black}}
\newcommand{\dsa}{\color{black}}
\newcommand{\jwa}{\color{black}}
\received{XXX}
\revised{YYY}
\accepted{ZZZ}

\submitjournal{AAA}

\shorttitle{POET \Rmnum{1}}

\shortauthors{An et al.}

\begin{document}
\begin{CJK*}{UTF8}{gbsn}

\title{Planetary Orbit Eccentricity Trends (POET). \Rmnum{1}. The Eccentricity-Metallicity Trend for Small Planets Revealed by the LAMOST-Gaia-Kepler Sample.
}

\correspondingauthor{Ji-Wei Xie}
\email{jwxie@nju.edu.cn}

\author{Dong-Sheng An (安东升)}
\affiliation{School of Astronomy and Space Science, Nanjing University, Nanjing 210023, People’s Republic of China}
\affiliation{Key Laboratory of Modern Astronomy and Astrophysics, Ministry of Education, Nanjing 210023, People’s Republic of China}

\author{Ji-Wei Xie (谢基伟)}
\affiliation{School of Astronomy and Space Science, Nanjing University, Nanjing 210023, People’s Republic of China}
\affiliation{Key Laboratory of Modern Astronomy and Astrophysics, Ministry of Education, Nanjing 210023, People’s Republic of China}

\author{Yuan-Zhe Dai (戴远哲)}
\affiliation{School of Astronomy and Space Science, Nanjing University, Nanjing 210023, People’s Republic of China}
\affiliation{Key Laboratory of Modern Astronomy and Astrophysics, Ministry of Education, Nanjing 210023, People’s Republic of China}

\author{Ji-Lin Zhou (周济林)}
\affiliation{School of Astronomy and Space Science, Nanjing University, Nanjing 210023, People’s Republic of China}
\affiliation{Key Laboratory of Modern Astronomy and Astrophysics, Ministry of Education, Nanjing 210023, People’s Republic of China}

\begin{abstract}

Orbital eccentricity is one of the basic planetary properties, whose distribution may shed light on the history of planet formation and evolution. 
Here, in a series of works on Planetary Orbit Eccentricity Trends (dubbed POET), we study the distribution of planetary eccentricities and their dependence on stellar/planetary properties.
In this paper, the first work of the POET series, we investigate whether and how the eccentricities of small planets depend on stellar metallicities (e.g., [Fe/H]).
Previous studies on giant planets have found a significant correlation between planetary eccentricities and their host metallicities. 
Nevertheless, whether such a correlation exists in small planets (e.g. super-Earth and sub-Neptune) remains unclear. 
Here, benefiting from the large and homogeneous LAMOST-Gaia-Kepler sample, we characterize the eccentricity distributions of 244 {\ds (286)} small planets in single {\ds (multiple)} transiting systems with the transit duration ratio method.
We confirm the eccentricity-metallicity trend that eccentricities of {\ds single} small planets increase with stellar metallicities. 
Interestingly, a similar trend between eccentricity and metallicity is also found in the radial velocity (RV) sample. 
{\jw We also found that the mutual inclination of multiple transiting systems increases with metallicity, which predicts a moderate eccentricity-metallicity rising trend.} 
Our results of the correlation between eccentricity {\jw (inclination)} and metallicity for small planet support the core accretion model for planet formation, and they could be footprints of self (and/or external) excitation processes during the history of planet formation and evolution.

\end{abstract}


\section{Introduction} \label{sec:intro}
Orbital eccentricity is one of the fundamental parameters in planetary dynamics, which provides crucial constraints on planet formation and evolution. 
Based on the fact that the solar system's planets have small orbital inclinations (mean $\sim 3^\circ$) and eccentricities (mean $\sim0.06$), Kant and Laplace in the 18th century put forward that the solar system formed from a nebula disk, laying the foundation for the modern theory of planet formation.

Since the discovery of 51 Pegasi b by \citet{1995Natur.378..355M}, the radial velocity (RV) method has been widely used to detect exoplanets and to measure their orbital eccentricities.
In contrast to the near circular orbits of solar system planets, exoplanets detected by RV are commonly found on eccentric orbits (mean eccentricity $\sim0.3$), which may imply that some violent dynamical processes, e.g., planet-planet scattering \citep{2008ApJ...686..580C,2008ApJ...686..621F,2010ApJ...711..772R} may occur in the history of exoplanet formation and evolution. 
Although the RV method plays an important role in measuring exoplanet eccentricity, it suffers from some notable biases and degeneracies which can cause considerable systematical uncertainties of eccentricity distributions \citep{2008ApJ...685..553S,2010ApJ...709..168A,2011MNRAS.410.1895Z}.
Furthermore, the majority of eccentricities measured by RV are limited for Jupiter sized giant planets.

After the launch of Kepler space telescope, the transit method began dominating the search of exoplanets. 
However, it is hard to measure eccentricities of individual planets via transit observations alone, except for under some special circumstances, such as giant planets with large eccentricities \citep{2012ApJ...756..122D}, systems with significant transit timing variations \citep[TTVs, e.g.,][]{2014ApJ...787...80H}, and planet host stars precisely characterized by asteroseismology \citep{2015ApJ...808..126V,2019AJ....157...61V}. 
Still, one can robustly constrain the eccentricity distribution of a sample of transiting planets from their transit duration ratio distribution provided the knowledge of the densities of their host stars \citep{2008ApJ...678.1407F}. 
Using the transit duration method and with the Large
Sky Area Multi-Object Fiber Spectroscopic Telescope \citep[LAMOST;][]{2012RAA....12.1197C} spectral characterization of the transit host stars, \citet{2016PNAS..11311431X} found an eccentricity dichotomy in Kepler planets, namely, Kepler singles are on eccentric orbits with mean eccentricity $\sim$ 0.3, whereas the multiples are on nearly circular and coplanar orbits similar to those of the solar system planets. 

Know thy star, know thy planet. 
Beside the numerous exoplanet discoveries in recent years, great progresses have also been made in characterizing the host stars of exoplanets with the help of various surveys of spectroscopy, e.g., Apache Point Observatory Galactic Evolution Experiment \citep[APOGEE;][]{2017AJ....154...94M}, California-Kepler
Survey \citep[CKS;][]{2017AJ....154..107P}, GALactic Archaeology with HERMES \citep[GALAH;][]{2012ASPC..458..421Z} and LAMOST \citep{2012RAA....12.1197C} and of astrometry, e.g., Gaia \citep{2018A&A...616A...1G}.

For example, benefiting from Gaia DR2 \citep{2018A&A...616A...2L}, \citet{2020AJ....159..280B}  produced the Gaia-Kepler Stellar Properties Catalog with uncertainties of stellar radius, stellar mass, stellar density and $\log{g}$ are $\thicksim $4\%, $\thicksim $7\%, $\thicksim $13\% and $\thicksim $0.05 dex respectively. 
In addition, LAMOST has observed more than one third of the Kepler targets without bias towards planet host stars \citep{2015ApJS..220...19D, 2018ApJS..238...30Z}, providing a homogeneous stellar characterization (e.g., $\log{g}$, RV, metallicity, etc.) in the Kepler field. 
Recently, by combining the data of LAMOST, Gaia and Kepler, \citet{2021AJ....162..100C} has built a catalog of kinematic properties (i.e., Galactic positions, velocities, and the relative membership probabilities among the thin disk, thick disk, Hercules stream, and the halo) as well as other basic stellar parameters for 35,835 Kepler stars, i.e., the LAMOST-Gaia-Kepler (LGK) sample, paving the road towards studying exoplanets in different Galactic environments and ages. 
Thanks to the advances both in exoplanet discovery and stellar characterization, we are allowed to carry out a series of work from here on, a project that we dub Planetary Orbit Eccentricity Trends (POET), aiming to reveal the dependence of orbital eccentricity on various stellar and planetary properties, and thus to provide new insights into planet formation and evolution.

Among various stellar properties, metallicity is a crucial factor in planet formation and evolution. 
On the one hand, metallicity can influence planetary occurrence rate. 
The occurrence of giant planets is strongly correlated to stellar metallicity (e.g., $\rm Fe/H$), i.e., the so called planet-metallicity relation, which has been well established \citep{2005ApJ...622.1102F} and is considered as one of key evidences to support the core accretion theory of planet formation \citep{2004ApJ...616..567I}. 
Recent studies have shown that the planet-metallicity relation may also extend to smaller planets \citep{2015AJ....149...14W,2019ApJ...873....8Z}, especially to those on short period orbits \citep{2016AJ....152..187M,2018PNAS..115..266D}.

On the other hand, metallicity may also influence the orbital architecture of a planetary system, e.g., eccentricity.
Both \citet{2013ApJ...767L..24D} and \citet{2018ApJ...856...37B} found that giant planets (e.g., Jovian planets) on eccentric orbits are preferentially residing with metal-rich stars.
However, the situation is still unclear for small planets, e.g., super-Earths and/or sub-Neptunes.
Using the asteroseismology sample, \citet{2019AJ....157...61V} studied the eccentricities of planets ($R_{p} < 6$ $R_{\oplus}$), and found no significant trend with stellar metallicity.
Nevertheless, using the CKS sample, \citet{2019AJ....157..198M} found tentative evidence that eccentric small planets prefer metal-rich stars.

Benefiting from the precise stellar parameters (e.g., mass, radius, and metallicity) provided by the LAMOST-Gaia survey \citep{2020AJ....159..280B, 2021AJ....162..100C}, and following the method from \citet{2016PNAS..11311431X}, here, in the first paper of the series of POET, we revisit the dependence of small planets' ($R_{p} < 4$ $R_{\oplus}$) eccentricities on stellar metallicities with the LGK sample \citep{2021AJ....162..100C}.
This paper is organized as follows.
In Section \ref{sec:data and sample selection}, we describe the  sample selection procedure. 
In Section \ref{sec:method}, we introduce the method of deriving eccentricity distribution. 
We present our results in Section \ref{sec:results} and discuss the implications of the results in Section \ref{sec:discussion and conclusion}. 
Finally, we give a summary in Section \ref{sec:summary}.

\section{sample selection} \label{sec:data and sample selection}

\begin{figure*}[ht!]
\includegraphics[width=1\textwidth]{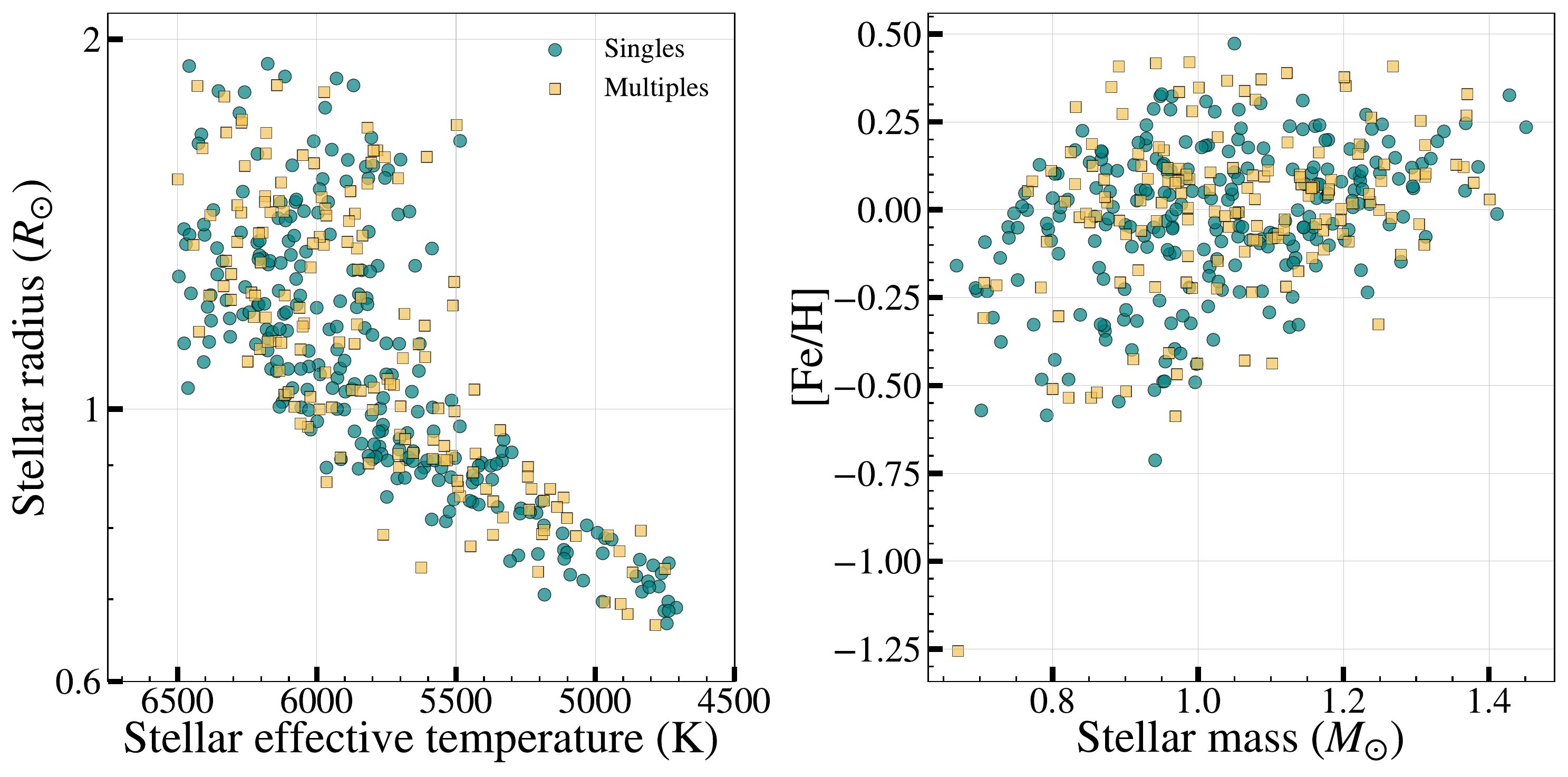}
\caption{Stellar properties in our sample. The left panel shows stellar effective temperature vs. stellar radius, and the right panel shows stellar mass vs. metallicity, i.e., [Fe/H]. \label{fig:stellar probity}}
\end{figure*}

We started our sample selection from the Kepler Data Release 25 \citep[DR25;][]{2018ApJS..235...38T,koidr25}, which has 6923 host stars with 8054 KOIs (Kepler Objects of Interest). 
First, we excluded false positives in our sample, after which 4034 confirmed/candidate planets were left around 3069 stars.
{\dsa To obtain a precise and homogeneous sample of metallicity, we then cross-matched the Kepler DR25 with LAMOST Data Release 8 (DR8). 
Note that we also cross-matched Kepler DR25 with LAMOST DR5 in the same way, and removed the stars whose [Fe/H] difference between DR5 and DR8 is greater than 3 $ \sigma $.
After this, 1409 planets in 1049 systems were left with a median metallicity uncertainty of $\thicksim$ 0.04 dex for the host stars.}
{\jwa This uncertainty of metallicity reflects only the internal uncertainty of LAMOST measurements (see Figure S1 of \citet{2016PNAS..11311431X}). 
For the systematic uncertainty, \citet{2016PNAS..11311431X} (see their Figure S2) found there is no significant offset but a larger dispersion, increasing the median of the total uncertainty of [Fe/H] to $\sim$0.1 dex.}
{\dsa In the following analyses, we will adopt a relatively large bin size of [Fe/H] ($\sim$0.15-0.6 dex) to reduce the effect of [Fe/H] uncertainty.}
Subsequently, we cross-matched the data with \citet{2020AJ....159..280B} for other stellar parameters, resulting in 1343 planets in 995 systems with median uncertainties of $\thicksim $4\%, $\thicksim $7\%, $\thicksim $0.05 dex in stellar radius, stellar mass, and $\log{g}$ respectively. 
To exclude the influence of potential binary stars, we adopted a cutoff of RUWE$<1.2$.\footnote{RUWE is the abbreviation of Gaia DR2's re-normalized unit-weight error computed by \citet{2020AJ....159..280B}. A large RUWE, e.g, RUWE$>$1.2, generally indicates the target is probably a binary.}
In addition, stars with GOF\footnote{GOF is the abbreviation of goodness-of-fit computed by \citet{2020AJ....159..280B}, a star with GOF $\leq$ 0.99 is an outlier with abnormal small errors.} $\leq$ 0.99 should be cautious \citep{2020AJ....159..280B}  and were excluded. 
We then adopted the following cuts, i.e., $\log{g}$ $>$ 4, $4700K < T_{\rm eff} < 6500K$ to focus on solar type main sequence stars. 
After all the above criteria on stars, there were left 899 planets in 638 systems.

We also applied the following criteria on planets to further refine the sample.
Following \citet{2011ApJ...736...19B}, we adopted a transit signal noise ratio cut SNR$>$7.1 to select reliable planet candidates.
Similar to \citet{2019AJ....157..198M}, here we also applied a cut on the uncertainty of the radius ratio of planet and star, i.e., relative error of $r\equiv \frac{R_{planet}}{R_{star}}<0.3$.
Following \citet{2018ApJS..235...38T}, we also selected KOIs with disposition score larger than 0.9 to have a reliable sample of planets.
Furthermore, as mentioned before the dependence of eccentricity on metallicity for gas giant planets have been relatively well established \citep{2013ApJ...767L..24D, 2018ApJ...856...37B} but not for small planets. 
Therefore, we only focus on small planets ($R_{p} < 4$ $R_{\oplus}$) here.
In addition, the orbit of planet with short period would be circularized via the tide between the host and the planet.
To avoid the influence of tide, we only considered planets with orbital period $P > 5$ days \citep[ e.g.,][]{2021ApJS..255....6D,2019AJ....157...61V}. 
Finally, we have 244 {\ds single transiting systems with 244  small planets} {\ds and 152 multiple transiting systems with 286 small planets} in our sample. {\dsn The data of the sample are provided in the Appendix \ref{sec:data} (Table \ref{tab:table2} and Table \ref{tab:table3}).}

\begin{figure}[ht!]
\centering
\includegraphics[width=.5\textwidth]{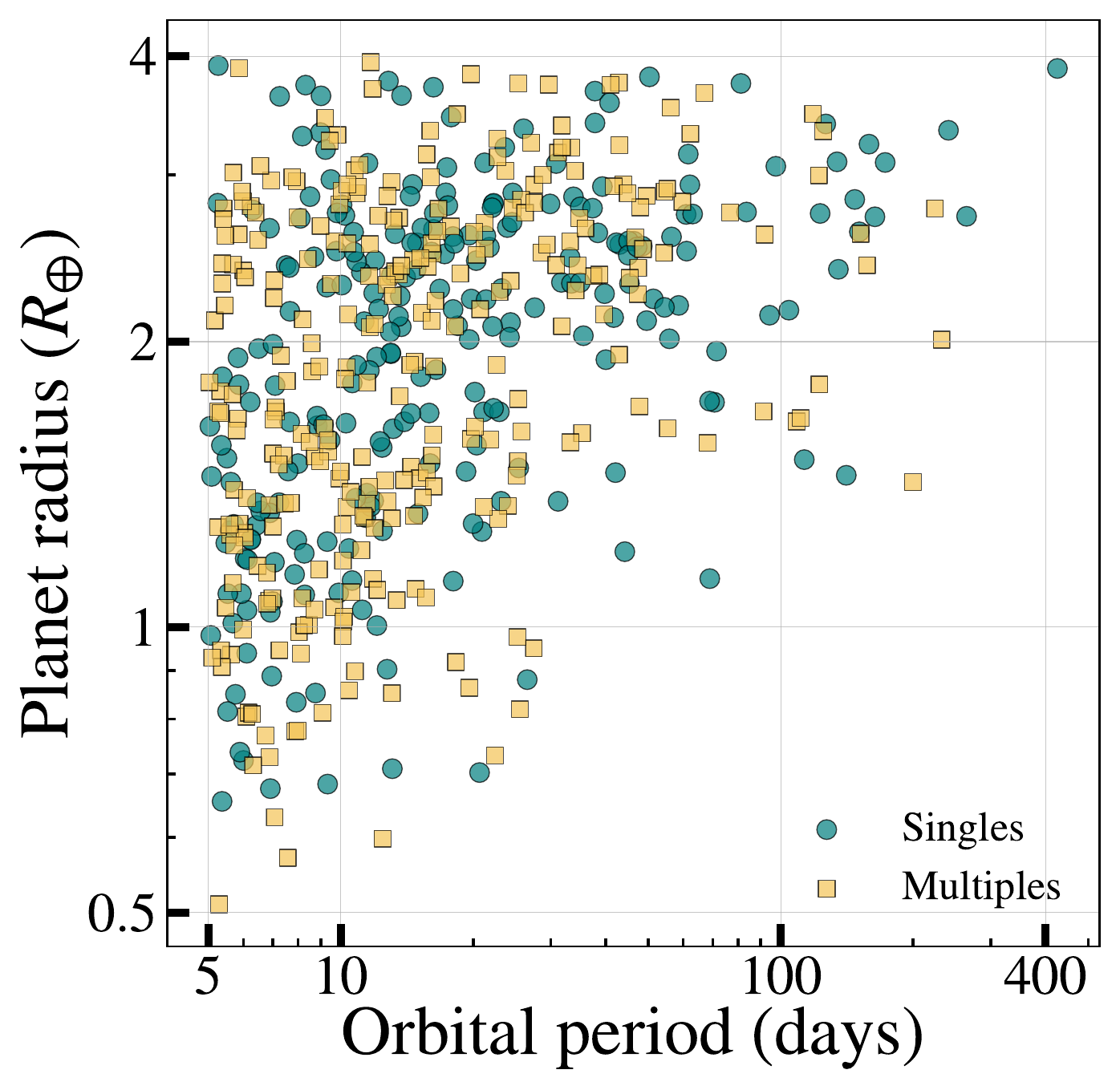}

\caption{Orbital period vs. radius of planets in our sample. \label{fig:planet probity}}
\end{figure}

Table \ref{tab:table1} is a summary of the sample selection process. 
Some basic properties of stars and planets in our sample are shown in Figure \ref{fig:stellar probity} and Figure \ref{fig:planet probity}, respectively. 
From the right panel of the Figure \ref{fig:stellar probity}, we can see a trend that stellar metallicity increases with stellar mass.
Thus, to study the relationship between eccentricity and metallicity, one should remove the potential effects of stellar mass (see Section \ref{sec:With control variable}).

\begin{deluxetable}{lchlDlc}
\tablenum{1}
\tablecaption{Summary of the sample selection\label{tab:table1}}
\tablewidth{0pt}
\tablehead{
\colhead{ } & \colhead{Host star} &
\multicolumn2c{Planet}\\
\colhead{ } & \colhead{(Number)}&
\multicolumn2c{(Number)}
}
\startdata
Kepler DR25  & 6923 &  & 8054 \\
Not false positive & 	3069 &  & 4034 \\
{\dsa Match with LAMOST} & 1049 &  & 1409 \\
Match with \citet{2020AJ....159..280B}  & 995 &  & 1343 \\
RUWE $\leq$ 1.2 & 862 &  & 	1166 \\
GOF $>$ 0.99 & 860 &  & 	1164 \\
Main sequence $(\log{g} > 4)$& 740 &  & 1024 \\
4700$K$ $<$ $T_{\rm eff}$ $<$ 6500K  & 638 &  & 899 \\
SNR (signal noise ratio) $>$ 7.1 & 632 &  & 891 \\
Relative error of $r\equiv \frac{R_{p}}{R_{*}}$ $<$ 0.3 & 618 &  & 875 \\
{\ds Single (Multiple)$^a$} & 450 (168) &  & 450 (424) \\
{\ds Disposition score $>$ 0.9} & 394 (166) &  & 394 (380) \\
{\ds $R_{p} < 4$ $R_{\oplus}$ } & 345 (164) &  & 345 (362) \\
{\ds Period $>$ 5 days } & 244 (152) &  & 244 (286) \\
\enddata
\begin{tablenotes}
    \footnotesize
    \item[1] $^a${\ds The error of transit duration, stellar radius, stellar mass are needed, thus one planet is removed for the absent of transit duration error.} 
\end{tablenotes}
\end{deluxetable}

\section{method} \label{sec:method}
We follow \citet{2016PNAS..11311431X} to derive eccentricity of our planet sample.
We briefly summarize it as the follows.
And the details of the method can be found in the supplementary of their paper (see also in \citet{2008ApJ...678.1407F} and \citet{2011ApJS..197....1M}).

The basic idea is to use the distribution of transit duration ratio ($TDR\equiv T/T_{0}$) to constrain the eccentricity distribution, where $T$ is the observed transit duration and $T_{0}$ is the reference transit duration which assumes the transit impact parameter $b=0$ and eccentricity $e=0$. 
For illustrative purpose, $TDR$ is a function of  $e$, $b$, and the argument of pericenter $\omega$, 
\begin{equation}
TDR\equiv\frac{T}{T_{0}} \thicksim \frac{\sqrt{(1-b^2)(1-e^2)}}{1+e\sin{\omega}}.
\label{equ:equ1}
\end{equation}
Since there are three unknowns ($e$, $b$ and $\omega$) in Equation \ref{equ:equ1}, one cannot solve the eccentricity individually. 
Nevertheless, by assuming reasonable distributions of $b$ and $\omega$, we can constrain the distribution of $e$ from the distribution of a sample of observed $T/T_{0}$.

In practice, we use a more precise formula to model the transit duration which is \citep{2010MNRAS.407..301K},
\begin{equation}
T_{\rm mod} \ = \ \frac{P}{\pi}\frac{(1-e^2)^{3/2}}{(1+e \sin{\omega})^2}\arcsin{(\frac{\sqrt{(1+r)^2-b^2}}{b\tan{i_{0}}})},
\label{equ:equ03}
\end{equation}  
where $P$ and $i_{0}$ are the orbital period and the inclination of the planet, respectively. 
And $i_{0}$ is related to $b$ via 
\begin{equation}
b=4.2P^{2/3}(\frac{\rho_{*}}{\rho_{\odot}})^{1/3}\frac{(1-e^2)}{1+e\sin{\omega}}\cos{i_{0}},
\label{equ:equ33}
\end{equation}  
where $\rho_{*}$ is the density of the host star.
$T_{0}$ is calculated as 
\begin{equation}
T_{0} \ = \ 13P^\frac{1}{3}(\frac{\rho_{*}}{\rho_{\odot}})^{-\frac{1}{3}}(1+r).
\label{equ:equ04}
\end{equation}
Then, we have the modeled transit duration ratio $TDR_{\rm mod}=T_{\rm mod}/T_{0}$, and the observed transit duration ratio $TDR_{\rm obs}=T_{\rm obs}/T_{0}$.
The distribution of the eccentricity is obtained by fitting $TDR_{\rm obs}$ with $TDR_{\rm mod}$.

Maximum likelihood method is used to conduct the eccentricity fitting process. 
The likelihood for a given model (assuming Rayleigh distribution function for planets with a mean eccentricity $\bar{e}$) to produce an observed TDR is \citep{2014ApJ...787...80H}
\begin{equation}
\mathcal{L}(TDR_{\rm obs}|\bar{e}, \bar{i})= \int\,P(TDR_{\rm mod}|\bar{e},\bar{i})*e^{\frac{-(TDR-TDR_{\rm obs})^2}{2\sigma_{TDR}^2}}dTDR.
\label{equ:equ01}
\end{equation}
where {\ds $\bar{i}$ is the mean mutual inclination (relevant only for multiple transiting systems) of the planets}. 
{\dsn And $\sigma_{\rm TDR}$ is the uncertainty of 
$TDR_{\rm obs}$, calculated by $\sigma_{TDR}=\sqrt{(\frac{\sigma_{T_{obs}}}{T_{obs}})^2+(\frac{\sigma_{\rho_{*}}}{3\rho_{*}})^2+(\frac{\sigma_{r}}{1+r})^2}$, where $\sigma_{T_{obs}}$, $\sigma_{\rho_{*}}$, and $\sigma_{r}$ are the uncertainties of $T_{obs}$, $\rho_{*}$, and $r$ respectively.} 
The first term, $P(TDR_{\rm mod}|\bar{e}, \bar{i})$ in Equation \ref{equ:equ01} is the probability that a modeled transit planet produces the corresponding TDR given the mean eccentricity $\bar{e}$ {\ds and the mean inclination $\bar{i}$}. 
And the second term reflects the consideration of uncertainty by assuming Gaussian error. 
{\dsn Here, we simply assume the distribution of $TDR_{obs}$ is symmetric about $TDR$. 
In Appendix \ref{sec:asymmetri}, we test the effect of asymmetric posterior distribution of $TDR_{obs}$.}

{\ds For single systems, $\bar{i}$ is not defined and there is only one fitting parameter, $\bar{e}$.} To obtain $P(TDR_{\rm mod}|\bar{e})$ {\ds for single planets}, we conduct the following simulations using the transit planets in our sample. 
First, for each planet, we assign it an orbital eccentricity $e$ from a Rayleigh distribution with a mean of $\bar{e}$, an argument of pericenter $\omega$ and an $\cos{i_{0}}$ from uniform distributions.  
We repeat this step if the assigned eccentricity is so high that let the planet hit the surface of the star, i.e., $a(1-e)<R_{*}+R_{p}$, where $a$ is the orbital semi-major axis.
Next, we calculate the impact parameter using Equation \ref{equ:equ33}. 
If the absolute value of the impact parameter is too high, i.e., $|b|>1+r$ to make a transit, then we go back to the above first step to restart the simulation. 
Then, we calculate the transit duration $T_{\rm mod}$ (Equation \ref{equ:equ03}) and $T_{0}$ (Equation \ref{equ:equ04}).
We also estimate the modeled transit signal noise ratio ($SNR_{\rm mod}$)  from the observed one ($SNR_{\rm obs}$), i.e., $SNR_{\rm mod}=\sqrt{T_{\rm mod}/T_{\rm obs}}*SNR_{\rm obs}$.
We set a simple criterion $SNR_{\rm mod} > 7.1$ to ensure the modeled transit planet to be detectable
\footnote{{\dsa Note, the criterion $SNR_{\rm mod} > 7.1$ adopted is rather simple. In reality, the detection efficiency function is complicated. For example, it increases gradually above 7.1, rather than a sharp transition as one would compute from an idealized model.}}.
Otherwise, we go back to the above first step to restart the simulation. 
We repeat above steps until each observed planet has 300 corresponding modeled transit duration ratio ($TDR_{\rm mod}$) in our simulated sample.
Finally, we use the Gaussian Kernel Density Estimation function to fit the distribution of all the simulated $TDR_{\rm mod}$ to obtain the probability density function, i.e, $P(TDR_{\rm mod}|\bar{e})$.
 
{\ds For multiple systems, the method to obtain $P(TDR_{\rm mod}|\bar{e}, \bar{i})$ is similar to single systems, except that the mutual inclinations in the system are correlated. 
We following \citet{2018ApJ...860..101Z},
\begin{equation}
\cos{i_{0}}=\cos{I}\cos{i}-\sin{I}\sin{i}\cos{{\dsa \phi}},
\label{equ:equzhu}
\end{equation} 
}
{\ds where $i_{0}$ is the inclination of the planet relative to the line of sight, $I$ is the invariable plane for the system ($\cos{I}$ is a uniform distribution), {\dsa $\phi$ is the phase angle and drawn from a uniform distribution independently, }$i$ is the inclination of the planet in the system relative to the invariable plane and drawn from a Rayleigh distribution with a mean inclination $\bar{i}$.}

We multiply the likelihood to produce each observed transit duration ratio to calculate the total likelihood $\mathcal{L}(TDR_{\rm obs}|\bar{e}, \bar{i})$.
{\ds For single systems, we only have one parameter, we calculate the total likelihood as a function of $\bar{e}$ , and fit it with an polynomial function. 
As shown in Figure \ref{fig:detail 3 bins 285}, the best fit $\bar{e}$ (where $\mathcal{L}(TDR_{\rm obs}|\bar{e})$ is the maximum) and the $1\sigma$ (68.3\%) confidence of interval are shown.}
{\ds For multiple systems, we map the total likelihood in the $\bar{e}$-$\bar{i}$ plane, and give the best fit of $\bar{e}$, $\bar{i}$ and corresponding $1\sigma$ (68.3\%) confidence of intervals (as shown in Figure \ref{fig:details for multiples(with ks)}).}

\section{results} \label{sec:results}
\begin{figure*}[ht!]
\centering
\plotone{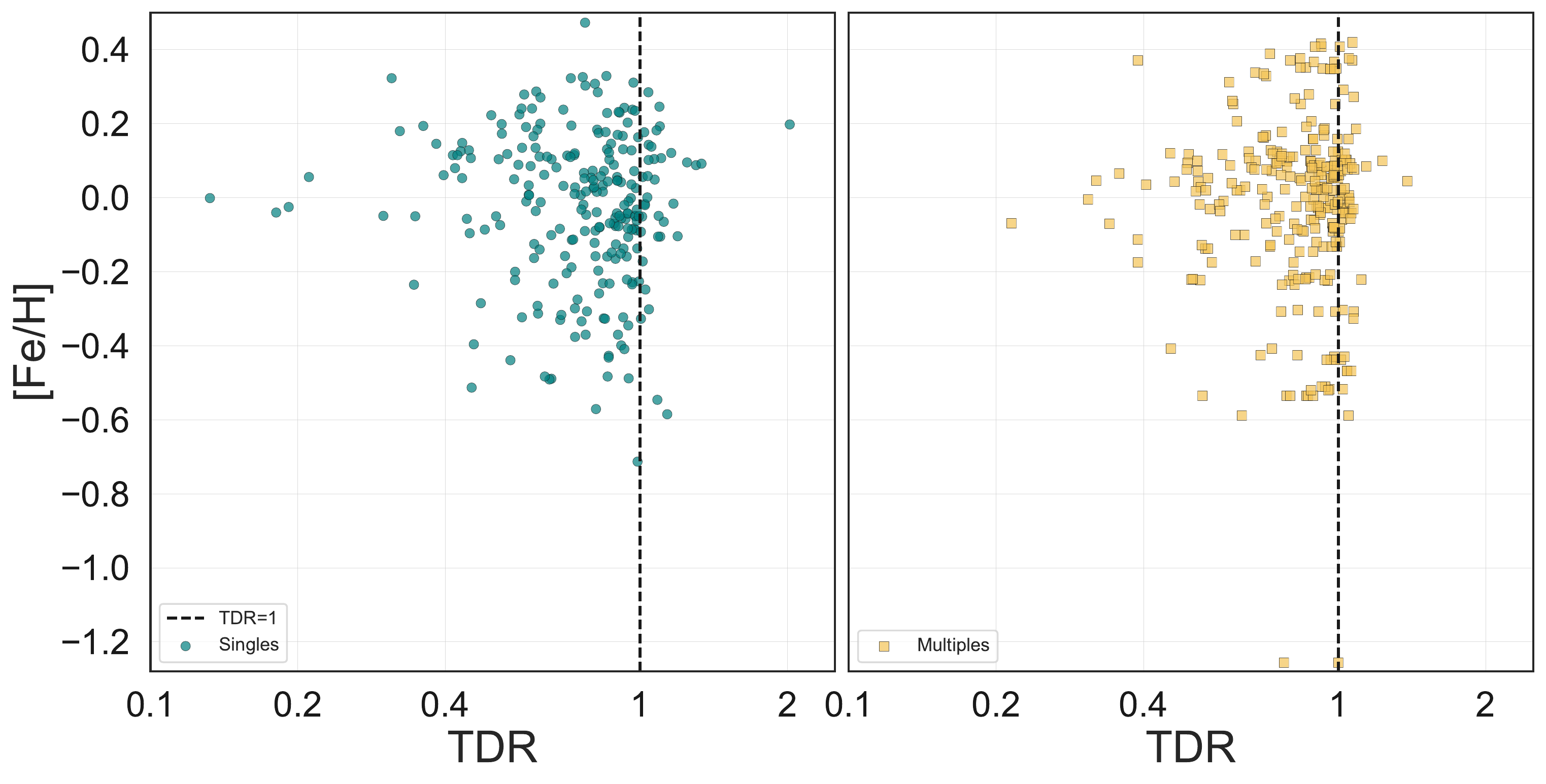}

\caption{{{\ds The metallicity ([Fe/H])- transit duration ratio (TDR) diagram for single (left panel) and multiple (right panel) transit systems.  }} \label{fig:3 bins before ks scatter}}
\end{figure*}

{\ds

Since the distribution of the transit duration ratio can reflect the distribution of eccentricity (Section \ref{sec:method}), we first plot a [Fe/H]-TDR diagram for singles (the left panel) and multiples (the right panel) respectively in Figure \ref{fig:3 bins before ks scatter} to have an intuitive feeling of transit duration ratio distribution as a function of metallicity. 
As can be seen, the TDR distribution for singles is apparently wider in metal-rich systems than in metal-poor systems, which qualitatively suggests larger planetary eccentricities associate with higher stellar metallicities. 
However, the TDR distribution for multiples is more concentrated around TDR=1 (indicating low eccentricities), and there is no significant dependence on metallicity.

}

\subsection{{\ds Single Transit Systems}}
\label{sec:results of singles}
\subsubsection{Without Parameter control: Metal Rich Stars Host High-e Planets}  \label{sec:Without control variable}

\begin{figure}[htbp]
\centering
\includegraphics[width=.5\textwidth]{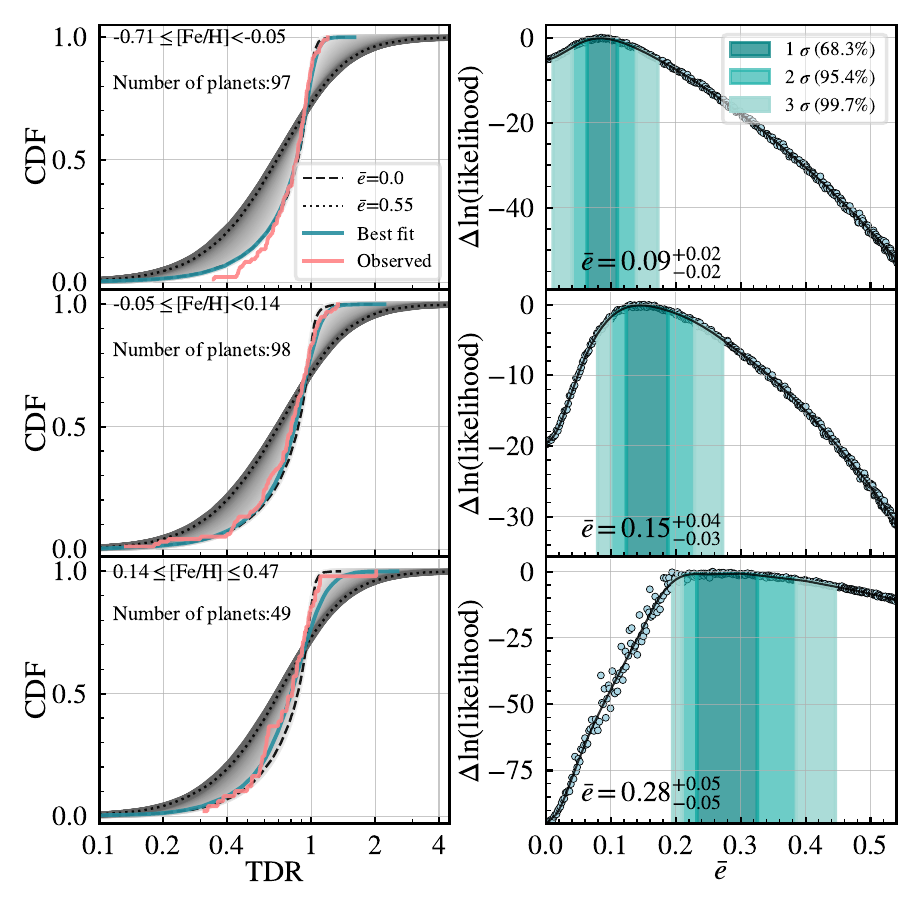}

\caption{Transit duration fitting for single transiting systems. Left panels: the cumulative distributions for modeled (gray) and observed (red) transit duration ratios (TDR). The range of metallicity and the number of planets for each sub-sample are also printed. Right panels: the relative likelihood of fitting the observed TDR distribution vs. the $\bar{e}$ assumed in the modeled TDR distributions, which gives the $1\ \sigma, \ 2\ \sigma,$ and $3\ \sigma $ confidence interval of $\bar{e}$. \label{fig:detail 3 bins 285}}
\end{figure}

\begin{figure*}
\centering

\plotone{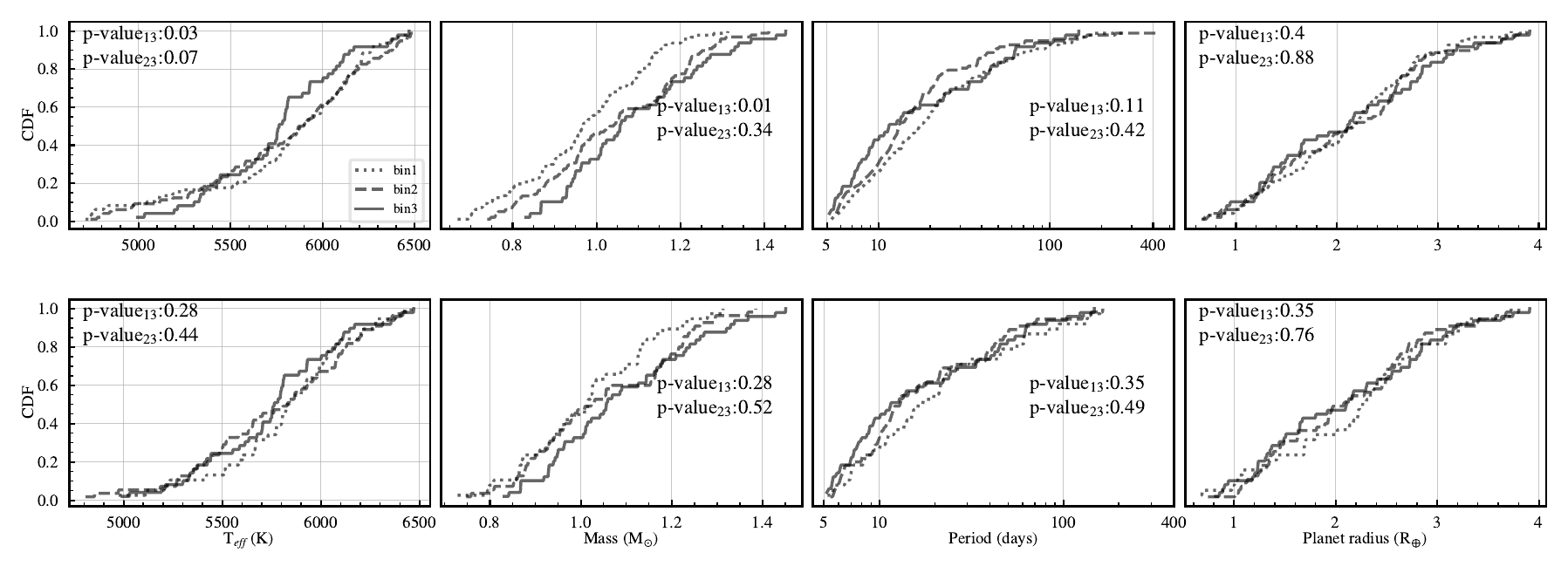}

\caption{Cumulative distributions for $T_{\rm eff}$, $M_{*}$, P, and $R_{p}$ in different sub-samples before (top four panels) and after (bottom four panels) parameter control, where bin1 (dotted line), bin2 (dashed line), and bin3 (solid line) represent the lowest, the intermediate, and the highest metallicity bin respectively. P-values of the Kolmogorov–Smirnov (KS) test between different 
bins are provided, where p-value$_{13}$ and p-value$_{23}$ represent the p-value of bin1 VS bin3 and bin2 VS bin3. \label{fig:ks before and after}}
\end{figure*}

In order to quantify the eccentricity distributions of different metallicities, we then divide the {\ds single} sample into three sub-samples according to [Fe/H].
Specifically, we first sort the whole {\ds single} sample in the order of [Fe/H].
We take the $\thicksim$ 20\% of systems at the highest [Fe/H] end as the metal-rich bin. 
And divide the rest into two bins in approximately equal size.
The reason that we artificially make the two latter bins larger is we can apply parameter control to further analysis (Section \ref{sec:With control variable}).

For each of the three sub-samples, we fit the transit duration ratio distribution to constrain the eccentricity distribution by following  the method as say in Section \ref{sec:method}. 
The results are shown in Figure \ref{fig:detail 3 bins 285}. 
As can be seen, $\bar{e} = 0.09_{-0.02}^{+0.02}$ in the lowest metallicity bin, $\bar{e} = 0.15_{-0.03}^{+0.04}$ in the intermediate metallicity bin, and $\bar{e} = 0.28_{-0.05}^{+0.05}$ in  the highest metallicity bin.
Above results quantitatively suggest that eccentricity increases with stellar metallicity for small single planets.

However, these results may be influenced by other parameters. 
For example, the above three sub-samples may differ in other parameters, e.g., stellar temperature, mass, planetary period, and radius (top panels of Figure \ref{fig:ks before and after}), which could also affect eccentricity distribution. 
In the next subsection we will perform a parameter control analysis to isolate the effect of metallicity on eccentricity.

\begin{figure}[htbp]
\centering
\includegraphics[width=.5\textwidth]{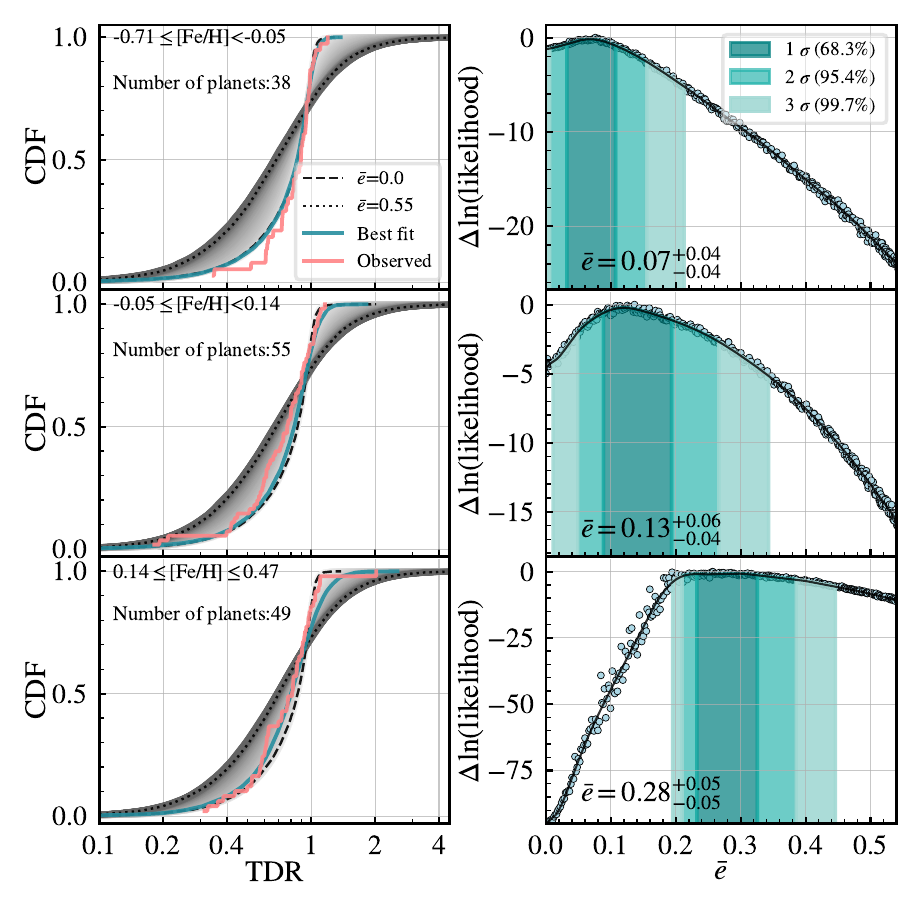}

\caption{Same as Figure \ref{fig:detail 3 bins 285}, but results for after parameter control on $T_{\rm eff}$, $M_{*}$, P, and $R_{p}$.  \label{fig:details for 3 bins(with ks)}}
\end{figure}

\subsubsection{With Parameter control: Minimize the Effects of other Stellar and Planetary Properties}\label{sec:With control variable}
In order to minimize the effects of other stellar and planetary properties, we control these parameters (stellar effective temperatures $T_{\rm eff}$, stellar masses $M_{*}$, planets' periods $P$, and planets' radii $R_{p}$) to let them have similar distributions in all the bins.
Specifically, for each system in the metal-rich bin  ($\thicksim$ 20\% of the whole sample) we search for the nearest two neighbors in the metal-poor and metal-intermediate bin respectively.
{\dsn We calculate the Euclidean distance (D) between planet systems as follows to find the nearest two neighbors. 
Specifically, $D=((k_1\frac{\Delta T_{\rm eff}}{\widetilde{\Delta T_{\rm eff}}})^2+(k_2\frac{\Delta M_{*}}{\widetilde{\Delta M_{*}}})^2+(k_3\frac{\Delta P}{\widetilde{\Delta P}})^2+(k_4\frac{\Delta R_{p}}{\widetilde{\Delta R_{p}}})^2)^{1/2}$
, where $\Delta T_{\rm eff}$, $\Delta M_{*}$, $\Delta P$, $\Delta R_{p}$ are the differences in stellar effective temperature, stellar mass, planetary period and planetary radius, and $\widetilde{\Delta T_{\rm eff}}$, $\widetilde{\Delta M_{*}}$, $\widetilde{\Delta P}$, $\widetilde{\Delta R_p}$ are the corresponding typical values {\jwa for scaling purpose, which are} calculated from the following procedure. 
For each of $N$ systems in the metal-rich bin, we calculate the $\Delta T_{\rm eff}$ between the system and systems in other bins, then find the smallest two $\Delta T_{\rm eff}$.
The $\widetilde{\Delta T_{\rm eff}}$ is calculated as the median of the $2\times N$ $\Delta T_{\rm eff}$.
And {\dsa $\widetilde{\Delta M_{*}}$, $\widetilde{\Delta P}$, $\widetilde{\Delta R_p}$} are calculated by following the same procedure. 
$k_1$, $k_2$, $k_3$, $k_4$ are four weighting coefficiencies. 
We tried different $k_1$, $k_2$, $k_3$, $k_4$ (range from 0.1 to 20) and performed the Kolmogorov–Smirnov (KS) test between the metal-rich bin and the selected neighbor systems in other bins for $T_{\rm eff}$, $M_{*}$, $P$, and $R_p$ to evaluate the goodness of finding the neighbors, and adopted the $k_1$, $k_2$, $k_3$, $k_4$ which lead to the highest p-value of KS test. 

{\dsa For example, before parameter control, we have 49, and 97 systems in metal-rich and metal-poor bin. 
We select the neighbors of the metal-rich bin in the metal-poor bin by following the steps below.
Step 1: we calculate the scaling parameters $\widetilde{\Delta T_{\rm eff}}=7.400K$, $\widetilde{\Delta M_{*}}=0.003M_{\odot}$, $\widetilde{\Delta P}=0.204$day, $\widetilde{\Delta R_p}=0.017R_{\oplus}$.
Step 2: for each system in the metal-rich bin, we select its two nearest neighbors in the metal-poor bin, i.e. the two systems with the smallest D values given a set of $k_1$, $k_2$, $k_3$, $k_4$. 
Step 3: we perform KS tests in the distributions of $T_{\rm eff}$, $M_{*}$, $P$, and $R_p$ between  the metal-rich bin and the selected metal-poor bin, and record the smallest p-value ($P_{KS}$) of the four KS tests.
Step 4: we repeat step 2 and step 3 for 10,000 times by adopting different sets of $k_1$, $k_2$, $k_3$, $k_4$ and choose the one with the highest $P_{KS}$ as the final result.
After the above steps and removing some duplicate systems, finally, we select 38 systems in the metal-poor bin as the neighbors of systems in the metal-rich bin.
}}

In Figure \ref{fig:ks before and after}, we show the distributions of these parameters before (top panels) and after (bottom panels) the parameter control process. 
Before the parameter control, bins of different metallicities differ significantly in the distributions of $T_{\rm eff}$ and $M_{*}$.
Metal-rich stars tend to have larger masses, which also can be seen in the right panel of Figure \ref{fig:stellar probity}.
After the parameter control, the metal-poor and metal-intermediate bins both have similar distributions in $T_{\rm eff}$, $M_{*}$, $P$, and $R_p$ as compared to the metal-rich bin (with all KS test p-values $\geq$ 0.28{\dsa , indicating that the two samples are likely to be drawn from the same distribution.}). 
{\dsa To further quantify how well the parameters have been controlled between different metallicity bins. We compare their median values and the corresponding 68.3$\%$ intervals. Specifically,
$T_{\rm eff}(K) = 5832_{-316}^{+282}$, $5828_{-464}^{+356}$, $5762_{-398}^{+338}$ after the parameter control for bin1, bin2 and bin3 in Figure \ref{fig:ks before and after}. The difference of the median, i.e.,  $T_{\rm eff}\sim70k$, between different bins is much smaller than the 68.3$\%$ interval and is even smaller than the typical measuring error of $\sim 100 k$. 
Similarly, 
$M_{*}(M_{\odot}) = 1.015_{-0.150}^{+0.132}$, $1.023_{-0.145}^{+0.202}$, $1.056_{-0.127}^{+0.201}$ after the parameter control for bin1, bin2 and bin3 in Figure \ref{fig:ks before and after}. 
The difference of the median, i.e.,  $M_{*}\sim0.04M_{\odot}(\sim 4\%)$, between different bins is much smaller than the 68.3$\%$ interval and is even smaller than the typical measuring error of $\sim 7\%$.
For plantary properties,
$P($days$) = 16.72_{-9.65}^{+44.23}$, $13.00_{-6.51}^{+31.86}$, $11.52_{-5.40}^{+39.94}$ after the parameter control for bin1, bin2 and bin3. 
The difference of the median, i.e.,  $P\sim 5.2$days, between different bins is much smaller than the 68.3$\%$ interval.
$R_p(R_{\oplus})=2.268_{-0.982}^{+0.816}$, $2.076_{-0.813}^{+0.744}$, $2.105_{-0.876}^{+0.872}$ after the parameter control for bin1, bin2 and bin3. 
The difference of the median, i.e.,  $R_p\sim 0.19 R_{\oplus}$, between different bins is much smaller than the 68.3$\%$ interval.}
{\dsa So far, $T_{\rm eff}$, $M_{*}$, $P$, and $R_p$ have been well controlled for different Fe/H bins, therefore, any significant eccentricity-Fe/H trend identified after the above parameter control should not affect by these controlled parameters.  




}

\begin{figure}[htbp]
\centering
\includegraphics[width=.5\textwidth]{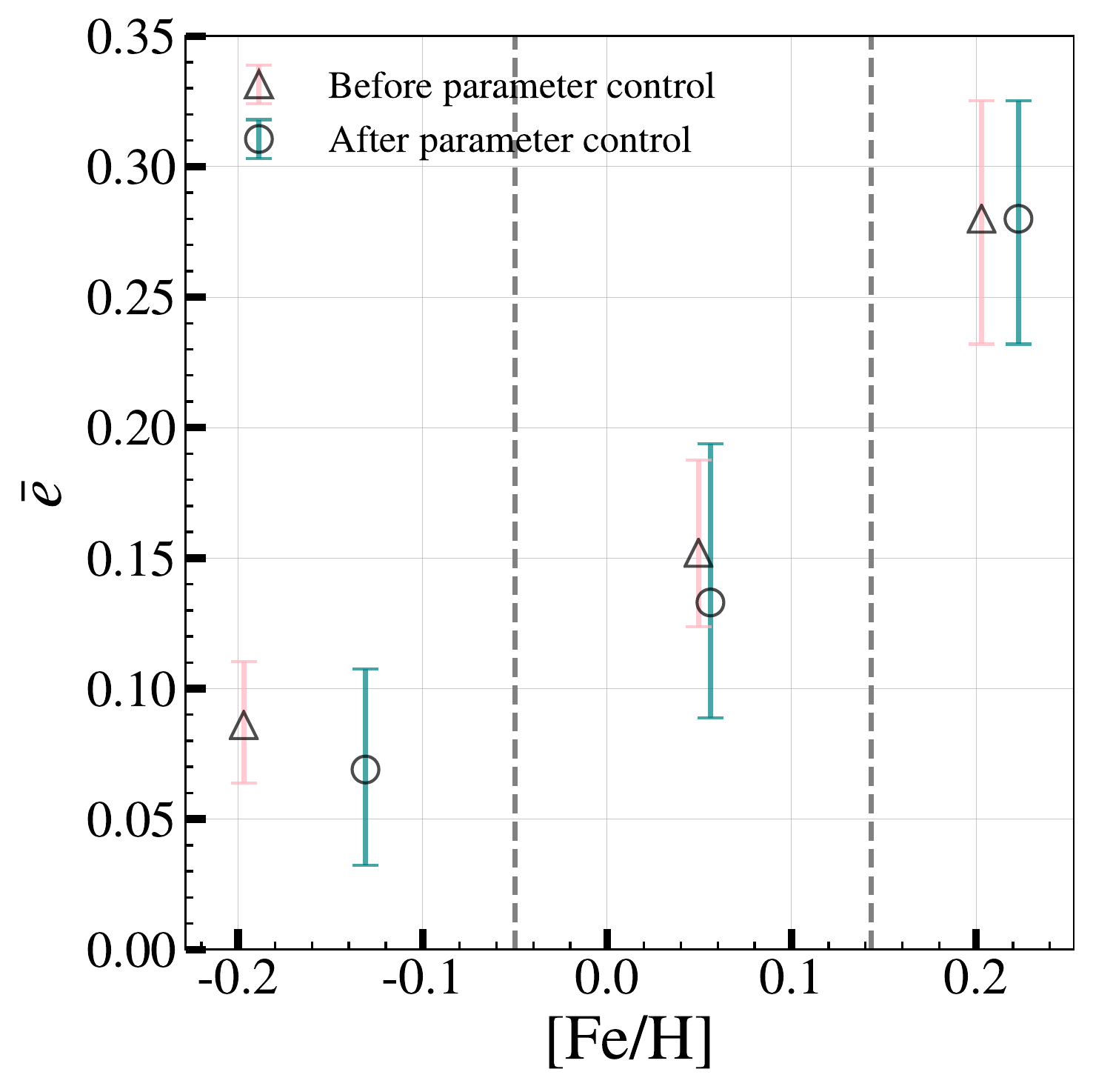}

\caption{Mean eccentricities ({\ds median $\bar{e}$ from} Section \ref{sec:Without control variable}, \ref{sec:With control variable}) vs. [Fe/H] (median [Fe/H] of each sub-sample), where error bars represent 1 $\sigma$ uncertainties of the mean eccentricity. And the gray dashed lines are the boundaries of each sub-sample. Boundaries between each sub-sample change less than 0.01 after parameter control, so we don't update the boundaries here. 
{\jw In addition, the result of the highest metallicity sub-sample before parameter control has been artificially moved to the left by 0.02 index to avoid overlap of symbols.}\label{fig:All outcomes for 3bins(with ks)}}
\end{figure}

Then for each controlled sub-sample, we preform the transit duration ratio distribution fitting as mentioned in Section \ref{sec:method} to obtain mean eccentricity. 
The fitting results are shown in Figure \ref{fig:details for 3 bins(with ks)}. 
The mean eccentricities of planets from the bin of lowest to highest in metallicity (from the top to the bottom panels) are $\bar{e} = 0.07_{-0.04}^{+0.04}$, $\bar{e} = 0.13_{-0.04}^{+0.06}$, and $\bar{e} = 0.28_{-0.05}^{+0.05}$ respectively.
Figure \ref{fig:All outcomes for 3bins(with ks)} compare the results before and after the parameter control.
The error bars become larger than before due to the reduction of number of planet in each sub-sample after parameter control. 
As can be seen, both show a similar trend that eccentricity increases with metallicity.
This may suggest eccentricity of small planet is not sensitive to stellar effective temperature and stellar mass (the main controlled parameters here).

\subsubsection{Effects of Binning} \label{sec:Divide into 4 bins}

\begin{figure}[htbp]
\centering
\includegraphics[width=.5\textwidth]{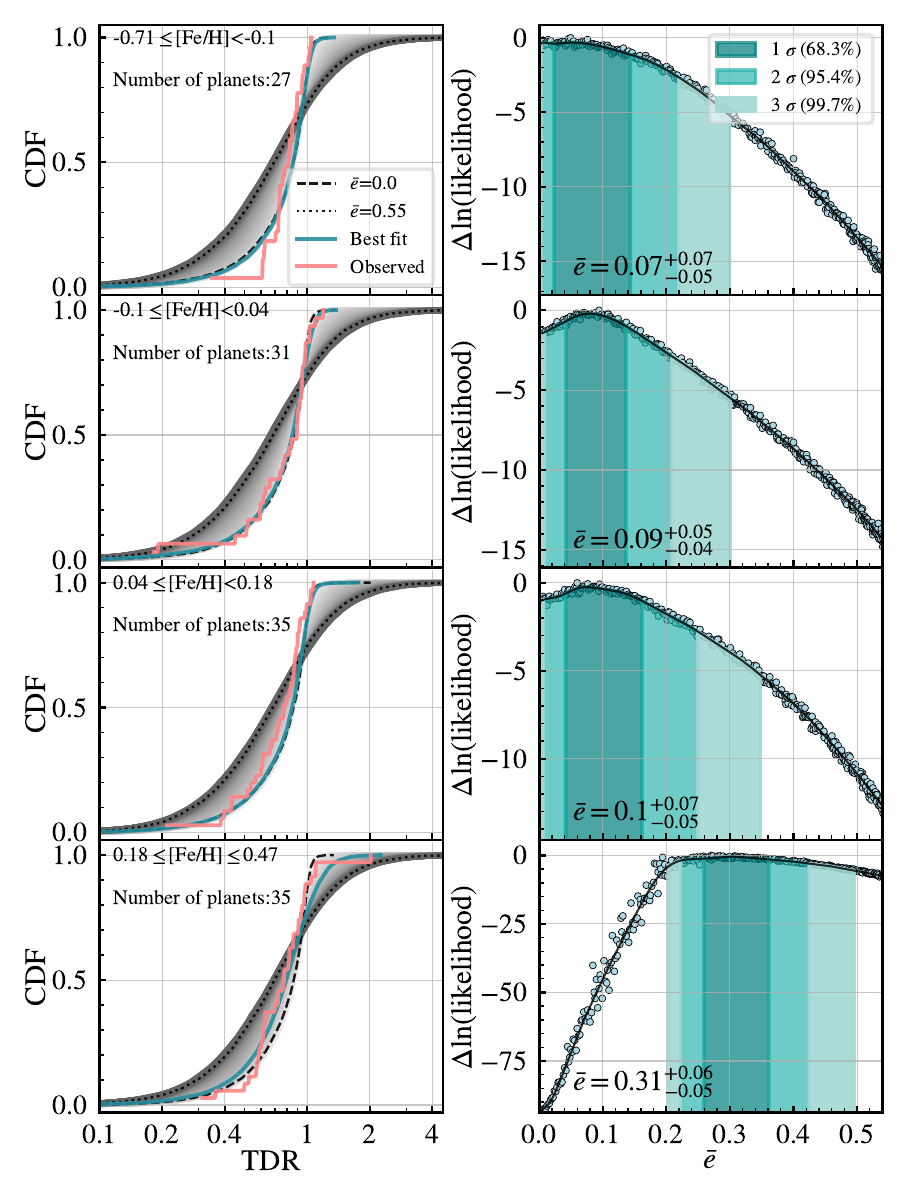}

\caption{Same as Figure \ref{fig:detail 3 bins 285}, but results for the four sub-
samples after parameter control on $T_{eff}$, $M_{*}$, P, and $R_{p}$. \label{fig:details for 4 bins(with ks)}}
\end{figure}

In our standard method, we set the size of the last bin as $\thicksim$ 20\% of the total sample, here we test whether our results are sensitive to the bin size .
Specifically, we consider two other cases with the last bin size as $\thicksim$ 15\% and $\thicksim$ 25\%, then preformed the same parameter control procedure and transit duration ratio fitting to derive eccentricity distribution.
For the case of $\thicksim$ 15\%, we find that $\bar{e} = 0.07_{-0.04}^{+0.04}$, $\bar{e} = 0.10_{-0.04}^{+0.06}$, and $\bar{e} = 0.31_{-0.05}^{+0.06}$ for the metal-poor, metal-intermediate, and metal-rich bins respectively.
For the case of $\thicksim$ 25\%, we find that $\bar{e} = 0.08_{-0.04}^{+0.05}$, $\bar{e} = 0.17_{-0.04}^{+0.05}$, and $\bar{e} = 0.27_{-0.04}^{+0.04}$ for the metal-poor, metal-intermediate, and metal-rich bins respectively.
As can be seen, the results of different bin sizes are consistent with each other (within 1 $\sigma$), and all show the same trend that planetary eccentricity increases with stellar metallicity. 

We also check the influence of the number of bins. 
We redivide all planets in our sample into four sub-samples, then perform the same parameter control procedure and transit duration ratio fitting to derive eccentricity distribution.
The results are shown in Figure \ref{fig:details for 4 bins(with ks)}: $\bar{e} = 0.07_{-0.05}^{+0.07}$, $0.09_{-0.04}^{+0.05}$, $0.10_{-0.05}^{+0.07}$, and $0.31_{-0.05}^{+0.06}$ for metallicity from low (top panel) to high (bottom panel) respectively.
These results also show that eccentricity increases with stellar metallicity, which is still consistent with the result in Section \ref{sec:With control variable} where we divide the sample into three sub-samples.

Therefore, we conclude that our results are not sensitive to the choice of bin size nor bin number.

\begin{figure*}[ht!]

\plotone{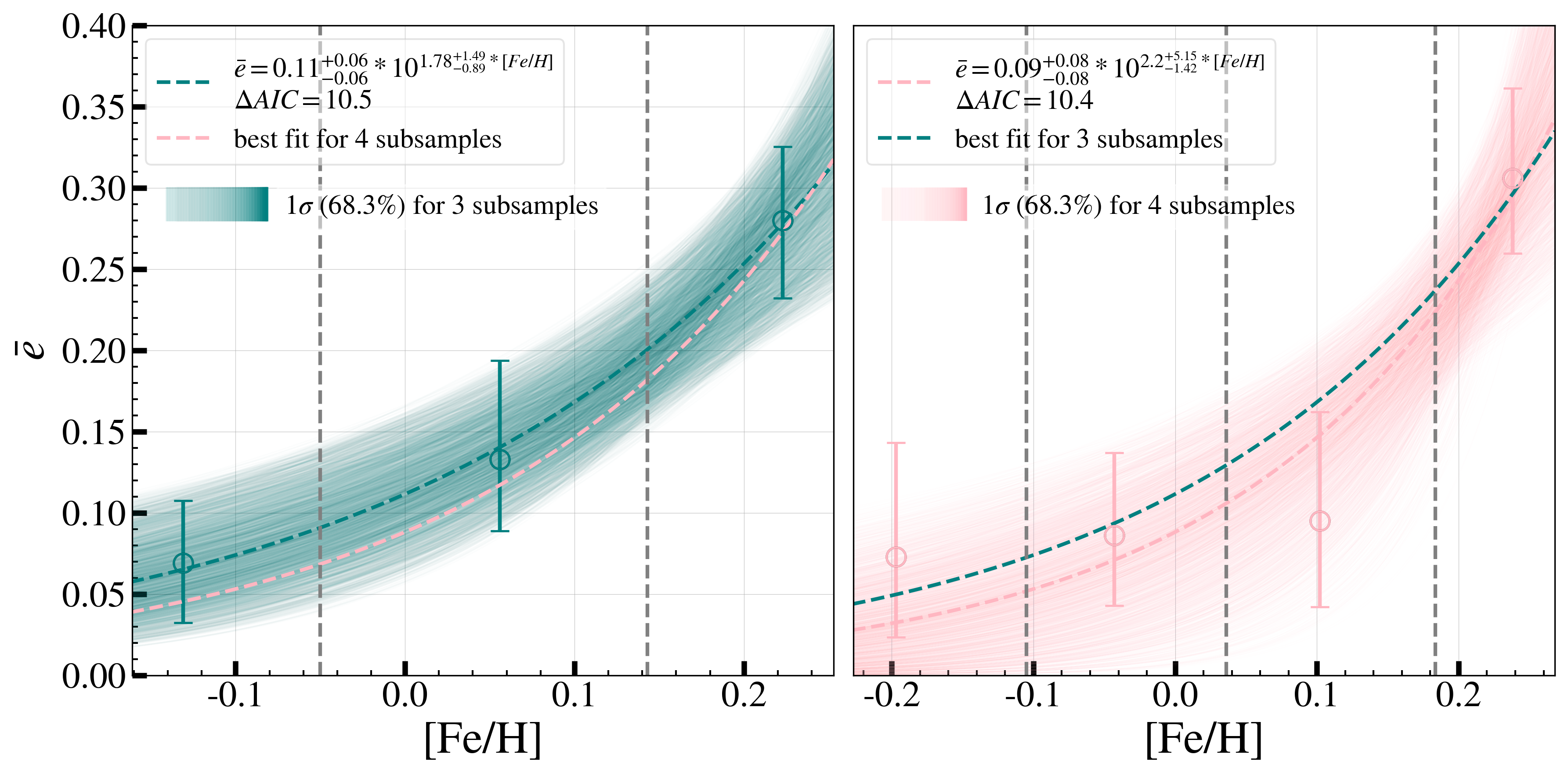}

\caption{Left panel: the fits for the three sub-samples, where the turquoise dashed line represents the exponential best fit. $\Delta AIC$ is AIC difference between the constant and exponential best fit. The light turquoise band is 1 $\sigma$ confidence interval for the exponential best fit. Right panel: same as the left panel, but shows the fits for the four sub-samples.  \label{fig:fit AIC}}
\end{figure*}

\subsubsection{Fit the Metallicity-Eccentricity Relation} \label{sec:Fit the results}

To quantitatively study the relationship between metallicity and eccentricity, here we fit our results by considering three models:  a constant model ($\rm {\ds\bar{e}}=constant$), a linear model ($\rm {\ds\bar{e}}=a*[Fe/H]+b$), and an exponential model ($\rm  {\ds\bar{e}}=a*10^{b*[Fe/H]}$) using least square method. 
In order to evaluate these models, we adopt the Akaike Information Criterion (AIC) \citep{1974ITAC...19..716A}.
Generally, a model with smaller AIC score is statistically better.
We calculate the AIC scores of best fits and corresponding parameters for the three models. 
For our standard case of three bins, {\ds $\rm AIC=14.5$, $\rm AIC=4.6$, and $\rm AIC=4.0$} for the constant, linear, and exponential models respectively. 
The exponential model is preferred with the smallest AIC.

In order to investigate the effect of the data uncertainty on the fitting results, we performed the following re-sampling and fitting analysis.
We re-sample {\dsa $\bar{e}$} according to the {\dsa {fitted probability distribution function of $\bar{e}$} (black curves in the } 
right panels of Figure \ref{fig:details for 3 bins(with ks)}) and re-fit the eccentricity-metallicity relation with the above three models. 
We repeat the re-sample and re-fit procedure 1,000,000 times and record the best fit parameters and the corresponding AIC in each time. 
For our standard case of three bins, we find that compared to the constant model, the linear model is preferred with smaller AIC in 966,932 times, and the exponential model is preferred in 978,835 times, corresponding to confidence levels of 96.7\% and 97.9\%.

Since the exponential model is preferred from the AIC analysis, we adopt it as our nominal model. 
{\dsa The best fit parameters are obtained by using the median $\bar{e}$ of each bins (Section \ref{sec:With control variable} and Section \ref{sec:Divide into 4 bins}) with the least square method.}
The 1 $\sigma$ confidence intervals of the model parameters are taken as $50\pm 34.1$ percentile of the {\dsa 1,000,000 times} re-sample fitting results.
The result for three sub-samples is 
\begin{equation}
{\ds\bar{e}=0.11_{-0.06}^{+0.06}*10^{1.78_{-0.89}^{+1.49}*[Fe/H]}}.
\label{equ:equ5}
\end{equation}
And the result for four sub-samples is
\begin{equation}
{\ds\bar{e}=0.09_{-0.08}^{+0.08}*10^{2.2_{-1.42}^{+5.15}*[Fe/H]}}.
\label{equ:equ6}
\end{equation}
These results are shown in Figure \ref{fig:fit AIC}.
As can be seen they are consistent within 1 $\sigma$.

To summary this section, we find the trend that the eccentricity increases with metallicity is robust and it is  best fit with an exponential model. 

\begin{figure}[htbp]
\centering
\includegraphics[width=.5\textwidth]{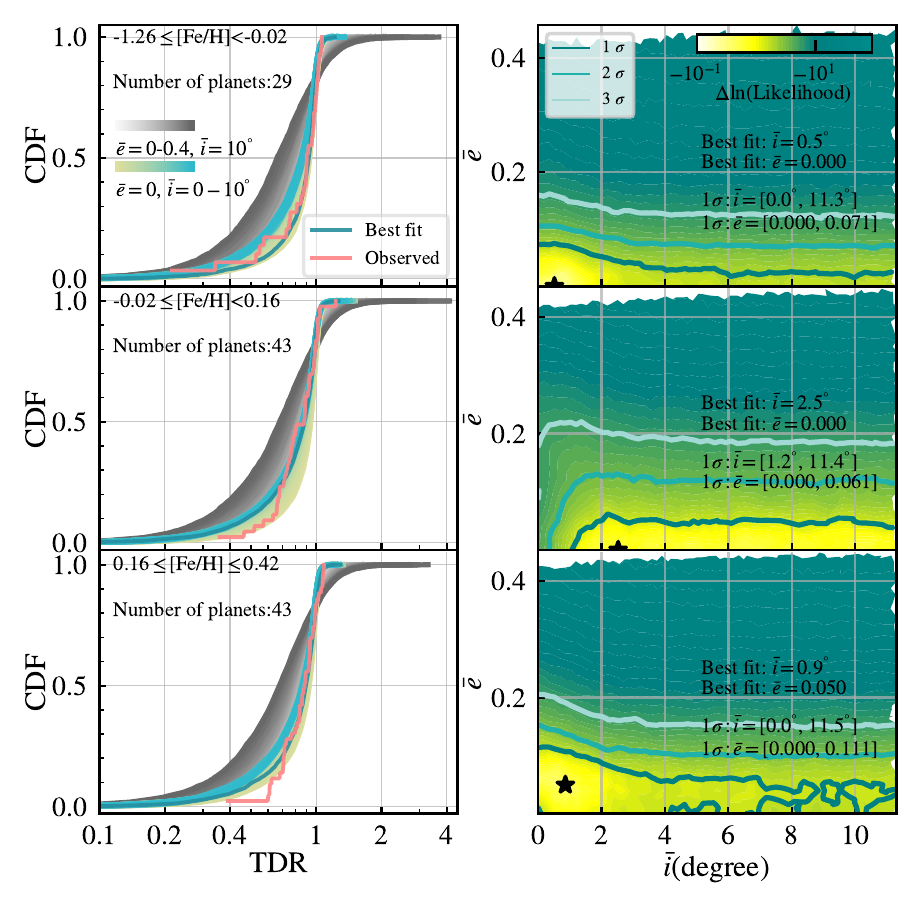}
\caption{{\ds Transit duration fitting for multiple transiting systems. Left panels: the cumulative distributions for modeled and observed (red) transit duration ratios. 
For comparison, simulations of $\bar{e}=0-0.4$ and $\bar{i}=10^{\circ}$ are shown in gray, simulations of $\bar{e}=0$ and $\bar{i}=0-10^{\circ}$ are shown in yellow to blue. 
Right panel: contours of relative likelihood in $\bar{e}$-$\bar{i}$ plane. 
The best fit of $\bar{e}$, $\bar{i}$ (also shown as black star) and corresponding confidence interval ($1\ \sigma, \ 2\ \sigma,$ and $3\ \sigma $) are also printed.}
\label{fig:details for multiples(with ks)}}
\end{figure}

\begin{figure}[htbp]
\centering

\includegraphics[width=.5\textwidth]{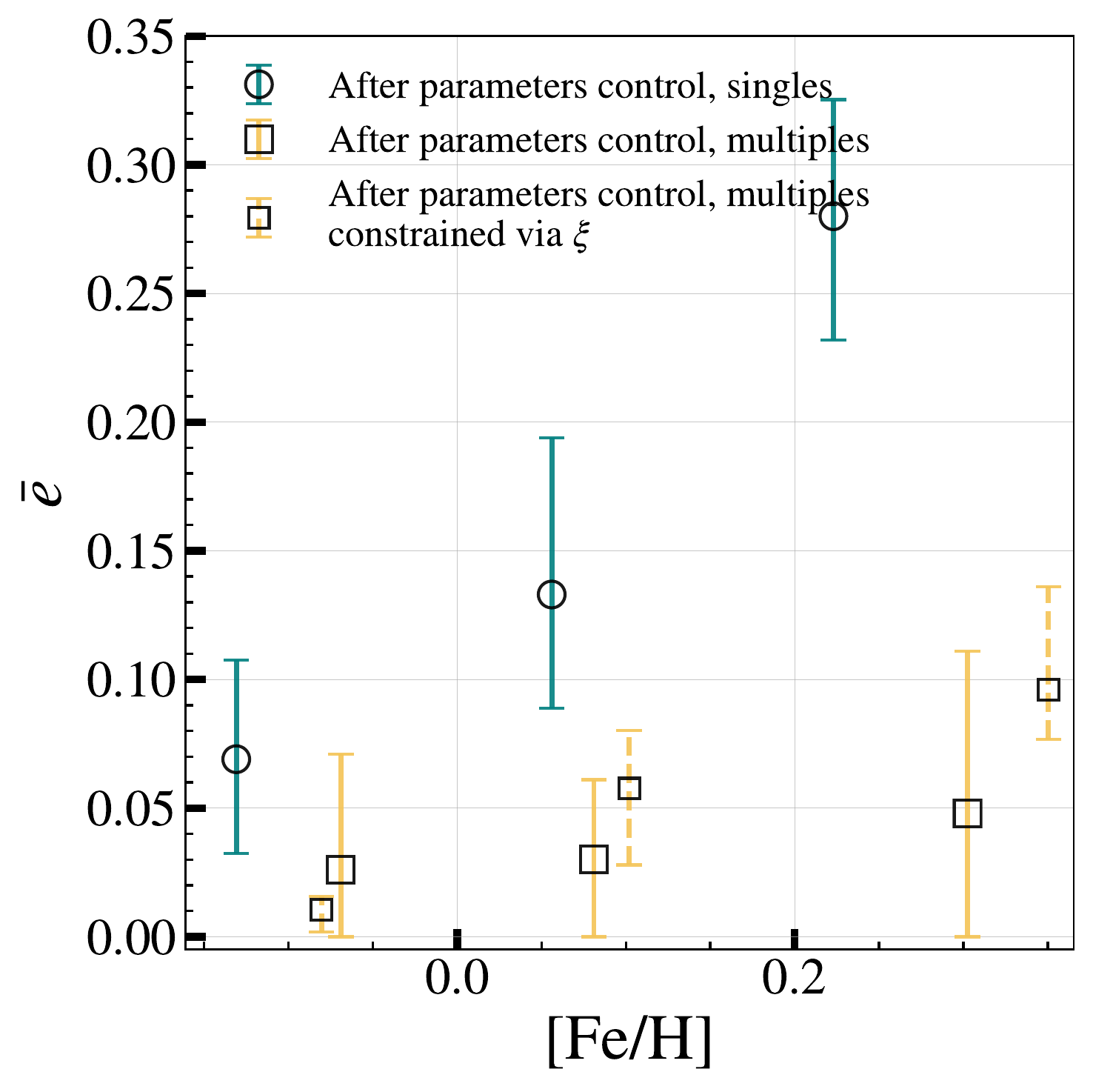}

\caption{{\jw Similar to Figure \ref{fig:All outcomes for 3bins(with ks)}, but showing mean eccentricities of singles vs. multiples under different metallicities. The solid error bars represent $\bar{e}$ constrained via TDR (Section {\ref{sec:results of singles}} and {\ref{sec:Metallicity-Eccentricity Trend}}), while the dashed error bars represent $\bar{e}$ calculated by assuming $\bar{e}=\bar{i}$ (see Section {\ref{sec:Comparison to previous study}}) and $\bar{i}$ is constrained by the distribution of mutual transit duration ratio, i.e., $\xi$ (Section {\ref{sec:Metallicity-Inclination Trend}}).
{\ds In addition, the dashed error bar of the medium metallicity has been artificially moved to right by 0.02 dex to avoid overlap of symbols.}
} \label{fig:singles vs multiples(with ks)}}
\end{figure}

{\ds

\subsection{Multiple Transit Systems} \label{sec:multiple planets}
\subsubsection{Metallicity-Eccentricity Trend} \label{sec:Metallicity-Eccentricity Trend}
Similar to Section \ref{sec:Without control variable} and Section \ref{sec:With control variable}, we divide the multiple planets' hosts into three sub-samples and preformed the same parameter control procedure. 
For each sub-sample of the multiples, we simulate the transit duration to constrain the mean eccentricity $\bar{e}$ and the mean inclination $\bar{i}$ via the method as mentioned in Section \ref{sec:method}. 
The results are shown in Figure \ref{fig:details for multiples(with ks)}.
As can be seen that the best fit of $\bar{e}$ are 0, 0 and 0.05 in the lowest metallicity bin, the intermediate metallicity bin and the highest metallicity bin respectively. 
If we use the median of $\bar{e}$, then $\bar{e}=0.026_{-0.026}^{+0.045}$, $0.030_{-0.030}^{+0.031}$ and $0.048_{-0.048}^{+0.063}$ from the bin of lowest to highest metallicity.
{\dsa We also fit the metallicity-e relation with the constant, linear and exponential model, and AIC=2.1, 4.0, 4.0 correspondingly. 
The constant model is preferred with the smallest AIC.}
Therefore, although the median eccentricity of multiples tend to increase slightly with metallicity, the trend is barely significant due to the relatively large uncertainties. 
As for inclination, the constrain is weak with much larger error bar compared to the eccentricity. 
We will revisit the {\jw metallicity-inclination} trend in Section \ref{sec:Metallicity-Inclination Trend}.

Figure \ref{fig:singles vs multiples(with ks)} shows the metallicity-eccentricity trend for singles and multiples.
While the singles shows a strong rising trend between eccentricity and metallicity, the trend is weaker for multiples given the relatively large uncertainties.
For all the metallicity bins, the multiples have smaller eccentricities than singles and the difference in eccentricity is larger for higher metallicity.

}

\begin{figure}[htbp]
\centering
\includegraphics[width=.5\textwidth]{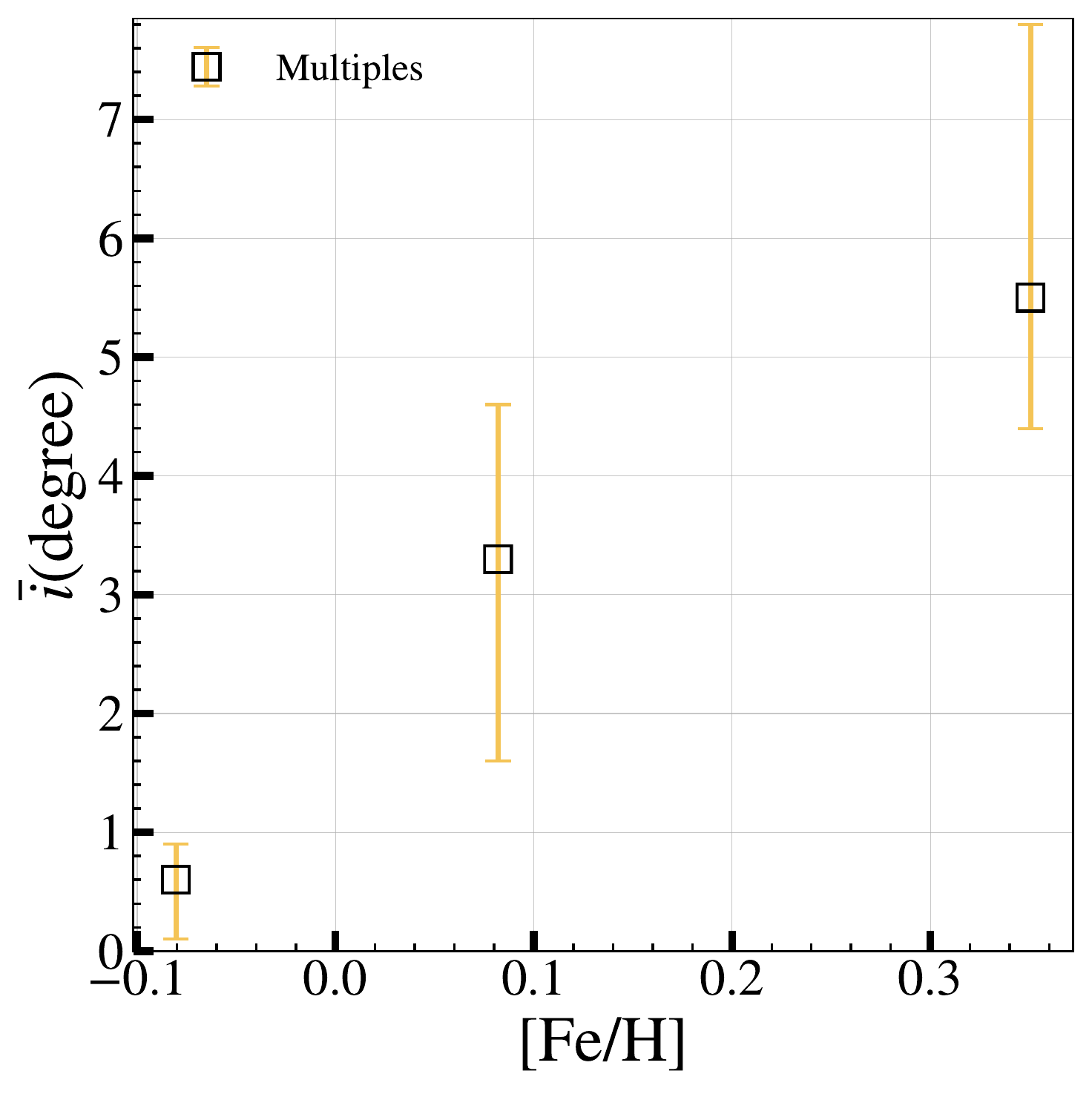}

\caption{ {\ds Mean mutual inclinations $\bar{i}$ of multiple transiting systems constrained by $\xi$ distribution (Section \ref{sec:Metallicity-Inclination Trend}) as a function of [Fe/H].
Each point
in the error bar represents the median of [Fe/H] and $\bar{i}$ in
each sub-sample} \label{fig:i vs feh}}
\end{figure}

{\ds
\subsubsection{Metallicity-Inclination Trend} \label{sec:Metallicity-Inclination Trend}

\begin{figure}[htbp]
\centering
\includegraphics[width=.5\textwidth]{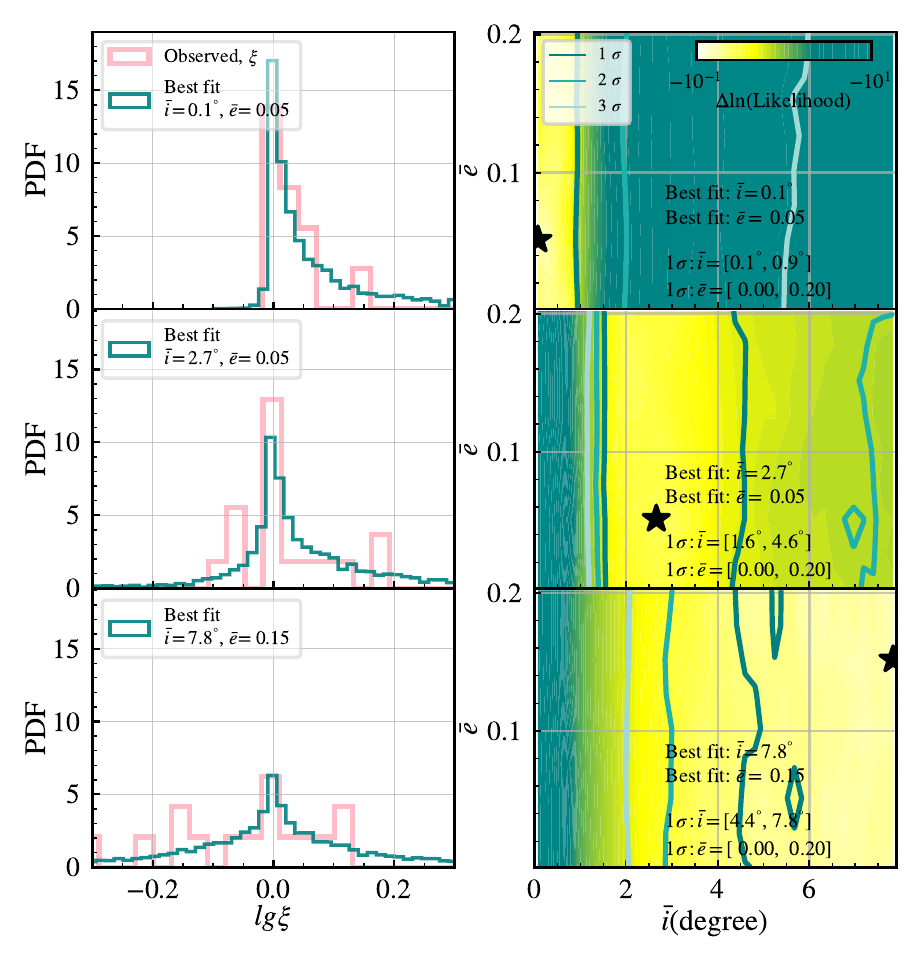}

\caption{ {\ds The mutual transit duration ratio ($\xi$) fitting for multiple transiting systems. Left panels: the observed $\xi$ distribution (red) and the best fit (turquoi). 
Right panels: the relative likelihood of fitting the observed $\xi$ distribution as a function of the $\bar{i}$ {\dsnn and $\bar{e}$} assumed in modeling the $\xi$ distributions, which gives the $1\ \sigma $ confidence interval of $\bar{i}$ {\dsnn and $\bar{e}$}.
The metallicty increases from the top to the bottom panels, i.e., [F/H]$<$-0.02, -0.02$\leq$[F/H]$<$0.16 and [F/H]$\geq$0.16 respectively.} \label{fig:xi fabrycky}}
\end{figure}

In addition to the normalized transit duration ratio (TDR, Equation \ref{equ:equ01}), there is a another metric, i.e., the mutual transit duration ratio ($\xi$), which is more sensitive to mutual inclination of multiple transiting systems. 
{\ds Following \citet{2014ApJ...790..146F},  $\xi$ is defined as
\begin{equation}
{\xi\equiv\frac{T_{in}/P_{in}^{1/3}}{T_{out}/P_{out}^{1/3}}},
\label{equ:equ_xi}
\end{equation}
where $T_{}$ and $P$ are the transit duration and orbital period, ``in'' and ``out'' represents the inner and the outer planets. 
Similar to \citet{2014ApJ...790..146F} and \citet{2018ApJ...860..101Z}, here we use $\xi$ to constrain the mean inclinations of the multiples (the same three sub-samples as in Section \ref{sec:Metallicity-Eccentricity Trend})\footnote{{\ds After the parameter control, some multiple systems just left one {\dsn small} planet, and thus these systems were excluded in the $\xi$ fitting process.}}. 
Following \citet{2018ApJ...860..101Z}, the likelihood of the simulated $\xi$ produce the observed $\xi_{j}$ under the mean inclination $\bar{i}$ {\dsnn and the mean eccentricity $\bar{e}$} is defined as,
\begin{equation}
\mathcal{L}=\prod_{j=1}^{N} \int\,P_{sim}(\ln\xi)*e^{\frac{-(\ln\xi_{j}-\ln\xi)^2}{2\sigma_{\ln\xi,j}^2}}d\ln\xi,
\label{equ:zhu ln}
\end{equation}
where $P_{sim}(\ln\xi)$ is the probability that the model produces the corresponding $\xi$ (in log scale) given the mean inclination $\bar{i}$ {\dsnn and the mean eccentricity $\bar{e}$}, $\xi_{j}$ is the observed $\xi$ for the $j$th planet pair, and $\sigma_{\ln\xi,j}$ is the corresponding uncertainty.
{\dsa Similar to modeling TDR as mentioned in section \ref{sec:method}, the modeling of $\ln\xi$ and thus the calculation of $P_{sim}(\ln\xi)$ also take into account the transit geometry effect and the detection efficiency by simply adopting a signal noise ratio SNR$_{mod}$ cut at 7.1.
The above likelihood function implicitly assumes that the $\ln\xi$ follows the Gaussian distribution.
In order to test if a Gaussian approximation can apply to $\ln\xi$, we randomly draw 100,000 $T_{in}$ and $T_{out}$ from Gaussian distributions given their reported values and errors, and take the value of $P_{in}$ and $P_{out}$ without errors (a good approximation) to calculate 100,000 $\ln\xi$ using Equation \ref{equ:equ_xi}. 
We find the distribution of $\ln\xi$ can be well fit by a Gaussian distribution.
}
Figure \ref{fig:xi fabrycky} and Figure \ref{fig:i vs feh} show the $\xi$ fitting results, where the mean mutual inclinations are constrained to be $\bar{i}=0.6_{-0.5}^{+0.3}$, $3.3_{-1.7}^{+1.3}$ and $5.5_{-1.1}^{+2.3}$ degrees from the bin of lowest (the top panel) to highest (the bottom panel) metallicity. 
Although with large uncertainties, the mean mutual inclination tends to increase with metallicity.

}
}


\section{discussions} \label{sec:discussion and conclusion}

\begin{figure*}[htbp]
\centering
\includegraphics[width=1\textwidth]{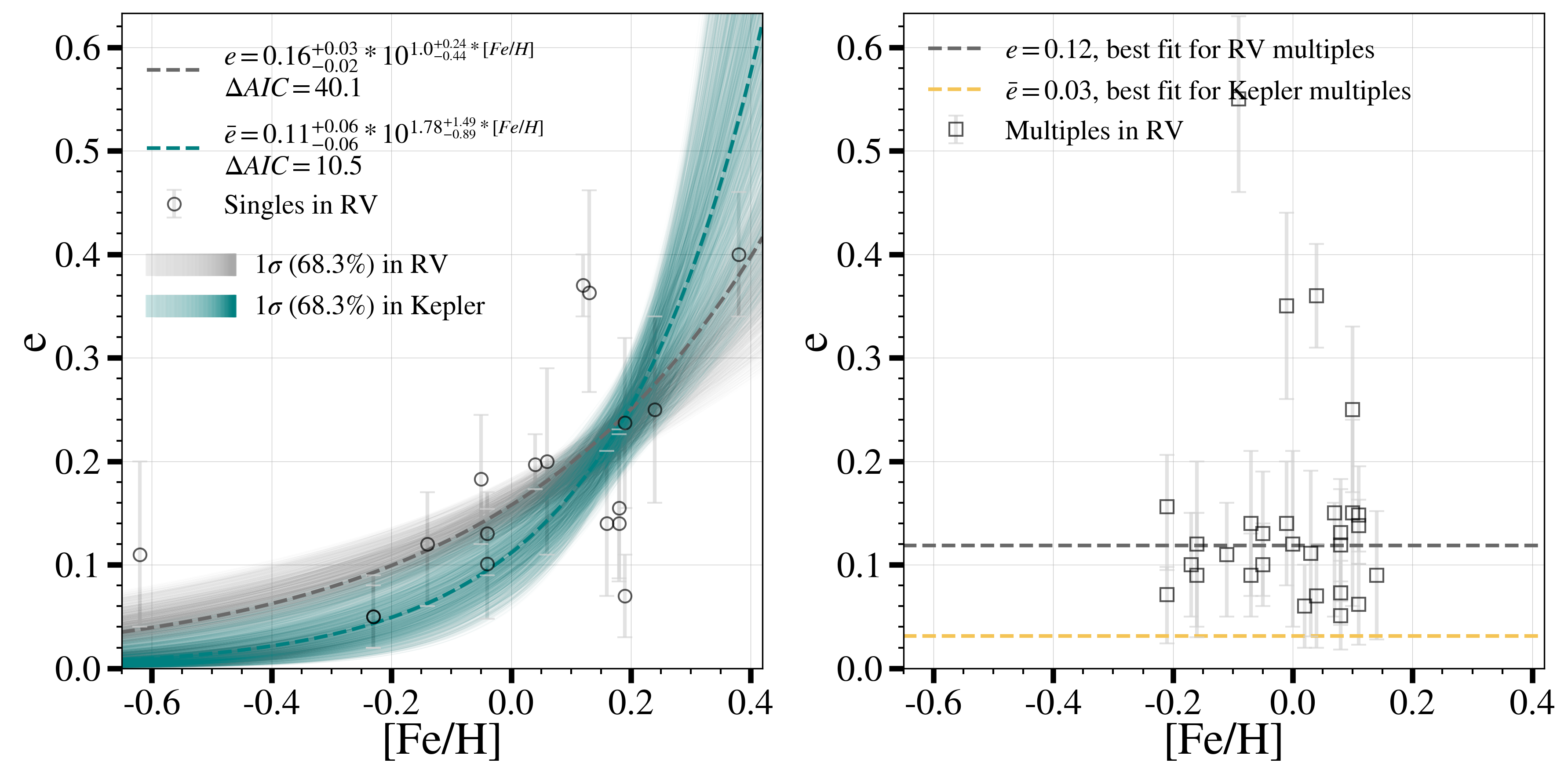}

\caption{{\ds Planet eccentricity (e) as a function of the stellar metallicity ([Fe/H]) for single (left panel) and multiple (right panel) systems in the RV sample. 
For the RV singles, the data prefer an exponential model and best fit is denoted by the gray dashed lines.
For comparison, we also plot the best fit for Kepler singles (Section \ref{sec:Fit the results}, the turquoise dashed lines).  
The colored bands represent 1 $\sigma$ confidence interval of the exponential best fit.
For the RV multiples, the data prefer a flat model, $e=0.12$ (gray dashed line). 
For comparison, we also plot the best fit of Kepler multiples ($\bar{e}=0.03$, the yellow dashed line).
}\label{fig:in RV}}

\end{figure*}

\subsection{Comparison to radial velocity planets} \label{sec:Comparison to radial velocity planets}

Our above results are based on the Kepler sample, here we compare them with that of the radial velocity (RV) sample.
Specifically, we download the RV sample from the NASA Exoplanet Archive \citep{ps}. 
To build a RV sample that is comparable to the Kepler sample, we select RV planets with the following criteria. 
First, we retain planets that have both reported eccentricities and host metallicities.
Then, we set conditions: $M\sin{i} < 32M_{\oplus}(\thicksim0.1M_{J})$ or $R_{p}<4R_{\oplus}$ to focus on small planets. 
Subsequently, we adopt an eccentricity quality cut, i.e., relative error of eccentricity $< 75\%$ and absolute error of eccentricity $< 0.1$. {\dsn And we also tried different relative error and absolute error cut, e.g., {\jw 100\% and 0.2}, and all the results are similar.}
{\ds In addition, similar to the Kepler sample, we only consider the systems with one (more than one) planet (planets) in $P<400$ days to build the RV single (multiple) sample. 
We also exclude planets with $P<5$ days to {\jw reduce} the influence of tide.
Finally, we have 18 (29) planets in the selected RV single (multiple) sample. {\dsn The data of these 18 (29) RV singles (multiples) are provided in the Appendix \ref{sec:data} (Table \ref{tab:table4} and Table \ref{tab:table5}).}
}

Figure \ref{fig:in RV} shows the distribution of the RV sample in the eccentricity-metallicity diagram.   
Apparently {\ds for singles in the left panel of Figure \ref{fig:in RV}}, there is also a trend that planetary eccentricity increases with stellar metallicity.
In order to qualify this trend, we preformed the same AIC analysis as in the Kepler sample (Section \ref{sec:Fit the results}). 
We find that {\ds $\rm AIC=106.2$, $\rm AIC=67.0$, and $\rm AIC=66.1$} for the constant, linear, and exponential models respectively. 
The exponential model is still preferred with the smallest AIC and the best fit (as print in Figure \ref{fig:in RV}) is consistent with the result of the Kepler {\ds singles} {\ds (Section \ref{sec:Fit the results})} within 1 $\sigma$. 

\subsection{Comparison to previous studies} \label{sec:Comparison to previous study}

\citet{2019AJ....157...61V} studied eccentricities of small planets using the asteroseismology sample and find no significant trend between {\ds single} planetary eccentricity and stellar metallicity.
Nevertheless, we note that their sample is small (only 30 planets with radius less than 6 $R_{\oplus}$) and the stellar metallicities are not homogeneous but coming from different literature.

\citet{2019AJ....157..198M} also studied the eccentricity of small planets by using the data from California-Kepler Survey (CKS). 
They identified 7 {\ds single} planets (6 with radius less than 4 $R_{\oplus}$) with high eccentricity and all of them are orbiting around metal-rich stars ([Fe/H]$>$0). 
Therefore, they tentatively conclude that small eccentric planets are preferentially found in high metallicity stars. 
In this study, with a large and homogeneous sample from the LAMOST-Gaia-Kepler catalog \citep{2021AJ....162..100C}, we confirm the trend that eccentricity increases with stellar metallicity for {\ds single} small planets. 
Encouragingly, a similar eccentricity-metallicity relation {\ds for singles} is also revealed by the RV sample (Figure \ref{fig:in RV}). 

{\jw 
\citet{2016PNAS..11311431X} shows that Kepler multiples and solar system objects follow a common relation ($\bar{e}\sim(1$–$2)\times\bar{i}$) between mean eccentricities and mutual inclinations. 
Given such a correlation between eccentricity and inclination {\dsa and the observed rising trend between inclination and metallcitity for multiple systems (Fig.\ref{fig:i vs feh}),  one may expect that the eccentricity of multiples should also increase with metallicity.}
However, our results show that, in multiple systems, the rising trend with metallicity is much weaker for  eccentricity than for inclination (Figure \ref{fig:singles vs multiples(with ks)} vs Figure \ref{fig:i vs feh}).
{\dsa One possible reason could be the much lower precision in measuring eccentricity than in measuring inclination for transiting systems. }
If we adopt a simple correlation $\bar{e}=\bar{i}$ and use the inclination-metallicity trend in Figure \ref{fig:i vs feh} to produce an expected eccentricity-metallicity trend (dashed bars in Figure \ref{fig:singles vs multiples(with ks)}), we find that current eccentricity measurements (solid bars in Figure \ref{fig:singles vs multiples(with ks)}) actually do not rule out such an expected trend considering the relatively large uncertainties.  

{\dsa
Another major finding of \citet{2016PNAS..11311431X} is that Kepler singles are on eccentric orbits with $\bar{e}\sim0.3$, whereas the multiples are on nearly circular($\bar{e}\sim0.04$).}
In this work, our measurement of $\bar{e}$ for singles is generally lower than that of \citet{2016PNAS..11311431X}.
This is probably because we exclude large planets (presumably have larger eccentricities than small planets) to focus on small planets here.
Nevertheless, in line with \citet{2016PNAS..11311431X}, our results also show that singles have larger eccentricities than multiples (Figure \ref{fig:singles vs multiples(with ks)}).
Furthermore, we find that the difference in eccentricity between singles and multiples increases with metallicity.
We will further discuss the implications of this result below.
}

\subsection{Implications to planet formation and evolution} \label{sec:Comparison to the simulation of planet formation}
The observed result that eccentricity increases with metallicity for small planet is not unexpected, instead it has important implications to planet formation and evolution.
According to the core accretion model for planet formation, planets form in proto-planetary disk through a bottom-up process from dust, planetesimals, planetary embryos and finally to full planets \citep{2004ApJ...616..567I}.
Generally, a higher stellar metallicity suggests a higher disk metallicity and thus more solid materials to form  planetesimals and more massive planets, which have stronger gravitational interactions to pump up larger orbital eccentricities.
Specifically, we expect two eccentricity exciting mechanisms as follows. 

On the one hand, eccentricities can be self excited among small planets themselves.
According to the N-body simulations by \citet{2016ApJ...832...34M}, the mean eccentricities of the planets increase from $\bar{e} \thicksim 0.06$ to $\thicksim 0.10$ when the total mass of the planetesimals in the disk increases from 7 $M_{\oplus}$ to 35 $M_{\oplus}$.
If the increase of the mass is caused by the increase of the metallicity completely, we can {\jw estimate} the corresponding metallicities ($Z$) by  $Z=-\lg(0.01M_{g}/M_{s})$, where $M_{s}$ and $M_{g}$ are the mass of solid and gas in the disk respectively \citep{2007MNRAS.378L...1G}. 
Here, we assume the total mass of the gas plus solid remain as a constant and mass of the disk to be 0.01 $M_{\odot}$ (\citet{10.1143/PTPS.70.35}, for the solar-like system) to estimate metallicity{\dsa , e.g., $M_{s}+M_{g}=0.01M_{\odot}$}.  
Therefore, the mass of solid increases from 7 $M_{\oplus}$ to 35 $M_{\oplus}$ corresponds to metallicity increases from [Fe/H]$\thicksim -0.68$ to [Fe/H]$\thicksim 0.03$.
Such an increase in metallicity leads to an eccentricity increase from $\thicksim 0.01$ to $\thicksim 0.12$ according Equation \ref{equ:equ5}. 
This result is comparable to the N-body  simulation result ($\bar{e}$ from $\thicksim 0.06$ to $\thicksim 0.10$) except at the metal-poorest end where Equation \ref{equ:equ5} underestimates the eccentricity somewhat.
Such a small inconsistency at the metal-poorest end is not unexpected because, in reality, a small eccentricity of a few percent is generally common for an essentially circular orbit. 
{\jw Furthermore, if taking an approximation that mutual inclination is correlated to eccentricity ($\bar{e}\sim1-2\times\bar{i}$, \citet{2016PNAS..11311431X}), then the inclination can be pumped up to $\bar{i}\sim0.05-0.1\sim2.9-5.7^\circ$ by self-excitation, which is comparable to our finding for Kepler multiples (the middle point of Figure \ref{fig:i vs feh}).} 
{\jw Based on the above simple estimation, we conclude that the self-excitation mechanism should, more or less, have played a role in producing the observed eccentricity (inclination)-metallicity trends.}

On the other hand, eccentricities of inner small planets can be excited and with the transiting multiplicity being reduced (producing more singles) at the same time, by perturbations of outer giant planets (e.g. \citet{2017AJ....153..210H}; \citet{2018MNRAS.478..197P}; \citet{2020MNRAS.498.5166P}).
Combining this with the well known metallicity-giant planet correlation \citep{2005ApJ...622.1102F}, one may naturally expect that a correlation between eccentricity and metallicity for small planets in single transiting systems.
Therefore, the outer-perturbed mechanism is complementary to the self-excitation mechanism, the combination of which is promising to fully explain the eccentricity-metallicity trend of singles.
{\jw Unfortunately, our sample of planets are detected by Kepler with the transit method, which strongly biases against long period planets, thus we could not test this mechanism in this work.
In the near future, the Gaia mission would find many long period giant planets, which combined with Kepler's short period small planets can further explore this scenario.
{\dsa In addition, RV surveys of planets found by TESS, K2 and PLATO could also help test the predictions of this scenario.}}

\section{summary} \label{sec:summary}

Since the discovery of 51 Pegasi b \citep{1995Natur.378..355M}, the number of exoplanets has increased dramatically. 
Furthermore, various surveys of spectroscopy and astrometry provide comprehensive characterizations for the host stars of exoplanets, allowing one to statistically study the relationship between stars and planets.
Here we start a project, Planetary Orbit Eccentricity Trends (POET) to investigate how orbital eccentricities of planets depend on various stellar/planetary properties.
In this work, the first paper of the POET series, we study the relationship between small planets' ($R_{p} < 4$ $R_{\oplus}$) eccentricities and stellar metallicities with the LAMOST-Gaia-Kepler sample \citep{2021AJ....162..100C}.

We found that, in single transiting systems, eccentricities of small planets increase with stellar metallicities (Figure \ref{fig:detail 3 bins 285}).
We excluded the influences of $T_{\rm eff}$, $M_{*}$, $P$, and $R_p$ on the eccentricity-metallicity trend by adopting a parameter control method to control these parameters (Figure \ref{fig:details for 3 bins(with ks)}, \ref{fig:All outcomes for 3bins(with ks)}). 
We also explored the effects of binning and found the eccentricity-metallicity trend is not sensitive to the size nor number of bins (Section \ref{sec:Divide into 4 bins} and Figure \ref{fig:details for 4 bins(with ks)}). 
Furthermore, we fitted  the eccentricity-metallicity trend and found it is best fitted with an exponential function (Section \ref{sec:Fit the results} and Equation \ref{equ:equ5}). 

{\jw In contrast, we found that, in multiple transiting systems, the eccentricity-metallicity rising trend is less clear. 
Although an inclination-metallcity trend is seen in multiples (Figure \ref{fig:i vs feh}) and it predicts a moderate eccentricity-metallcity rising trend as well, such an eccentricity-metallcity rising trend can neither be established nor be ruled out given the relatively large uncertainty in measuring eccentricity (Figure \ref{fig:singles vs multiples(with ks)}).
}

We then compared our results with the data from RV, and found they are consistent within 1 $\sigma$ (Figure \ref{fig:in RV}). 
We also compared our results with \citet{2019AJ....157...61V} and \citet{2019AJ....157..198M} in Section \ref{sec:Comparison to previous study}, where we emphasized the importance to have large and homogeneous stellar parameters when studying the relation between stars and planets. 
{\jw 
{\dsa Our results have shown that the difference of the mean eccentricity between singles and multiples increases with metallicity.}
Finally, we discussed the implication of the eccentricity-metallicity trend on planet formation and evolution (Section \ref{sec:Comparison to the simulation of planet formation}).
We identified two mechanisms (self-excitation and external excitation) that could potentially explain the observed eccentricity-metallicity trend. 
Future studies of both simulations and observations on a larger sample will further test them.

}

\acknowledgments
{\ds The authors thank the referee for the constructive comments and suggestions. }
{\jw The authors also thank Fei Dai, Jia-Yi Yang, Di-Chang Chen, Rui-Sheng Zhang, Ke-Ting Shin, and Tong Bao for helpful discussions.}
This work is supported by the National Key R$\&$D Program of China (No. 2019YFA0405100) and the National Natural Science Foundation of China (NSFC; grant Nos. 12150009, 11933001, and 12273011). 
J.-W.X. also acknowledges the support from the National Youth Talent Support Program and the Distinguish Youth Foundation of Jiangsu Scientific Committee (BK20190005).
This work has included data from Guoshoujing Telescope. 
Guoshoujing Telescope (the Large Sky Area Multi-Object Fiber Spectroscopic Telescope LAMOST) is a National Major Scientific Project built by the Chinese Academy of Sciences.
Funding for the project has been provided by the National Development and Reform Commission. LAMOST is operated and managed by the National Astronomical Observatories, Chinese Academy of Sciences.

\begin{figure}[htbp]
\centering
\includegraphics[width=.5\textwidth]{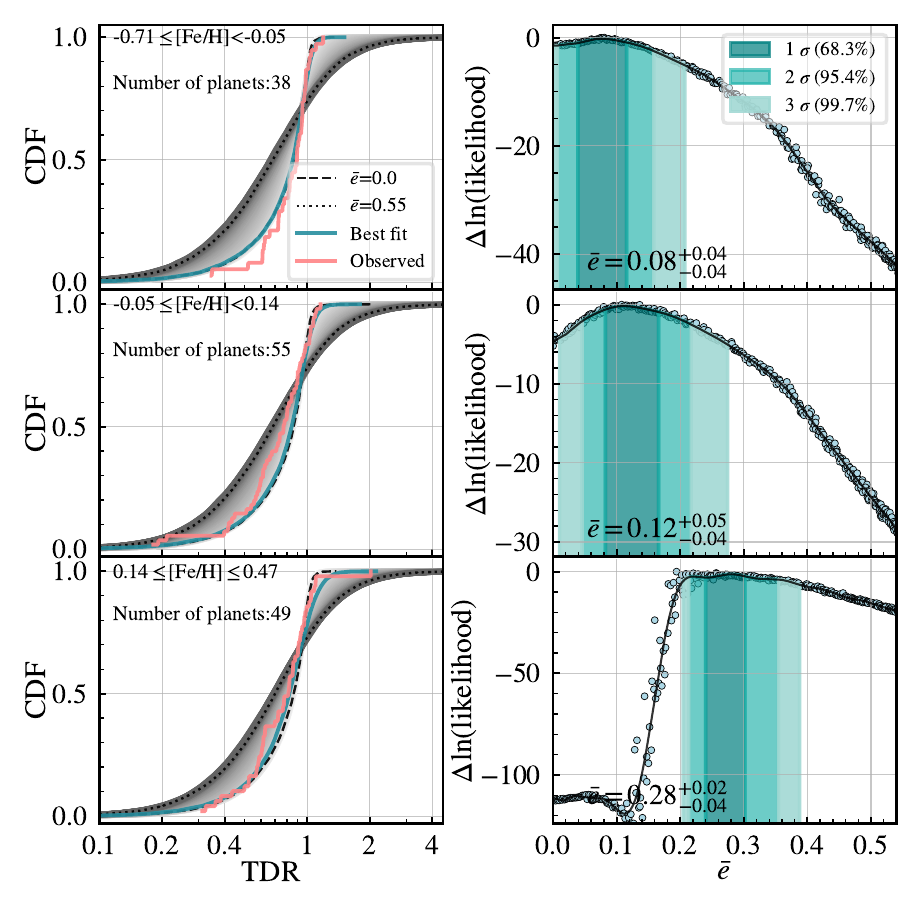}

\caption{Same as Figure \ref{fig:details for 3 bins(with ks)}, but results for the consideration of asymmetry density. \label{fig:details for 3 bins(asymmetri)}}
\end{figure}

\newpage
\appendix
{\dsn
\section{The effect of the asymmetric distribution of \texorpdfstring{$TDR_{obs}$}{Lg} \label{sec:asymmetri}}

In the main text of this paper, we have assumed the distribution of the $TDR_{obs}$ is symmetric about $TDR$ in the likelihood function (Equation \ref{equ:equ01}). 
However, the percentage of stars below the zero age main sequence (ZAMS) is small for Kepler stars, which will lead to an asymmetric distribution of $TDR_{obs}$. 
Specifically, to test the effect of the asymmetry of stellar density, we modify the uncertainty term in likelihood function as, 
\begin{equation}
\sigma_{TDR}=\sqrt{(\frac{\sigma_{T_{obs}}}{T_{obs}})^2+K(\frac{\sigma_{\rho_{*}}}{3\rho_{*}})^2+(\frac{\sigma_{r}}{1+r})^2}
\begin{cases}
K =0& \text{ $TDR< TDR_{obs} $ } \\
K=1& \text{ $ else$ }
\end{cases}
\label{equ:asymmetri}
\end{equation}
where $K$ is a coefficient to take into account the asymmetry of stellar density.
Here, we consider an extreme case, i.e., $K=0$ for $TDR< TDR_{obs}$ and $K=1$ for $TDR\geq TDR_{obs}$. 
} 
{\dsa This is equivalent to adopting a mixture of two half-Normal distributions with different widths for each observed $TDR_{obs}$. }
{\dsn
Figure \ref{fig:details for 3 bins(asymmetri)} shows the results for the 3 sub-samples (same sub-samples as in Section \ref{sec:With control variable}) by using the above modified uncertainty term. 
As can be seen, the mean eccentricities are  $\bar{e} = 0.08_{-0.04}^{+0.04}$, $\bar{e} = 0.12_{-0.04}^{+0.05}$, and $\bar{e} = 0.28_{-0.04}^{+0.02}$ from the lowest [Fe/H] bin to the highest [Fe/H] bin respectively. These results are very close to our norminal results in Section \ref{sec:With control variable}, where we assumed a symmetric $TDR_{obs}$. 
Therefore, we conclude that the asymmetric distribution of $TDR_{obs}$ has little effect on our results.
}
{\dsn
\section{Data of the planet systems used in this paper \label{sec:data}}
}
{\dsn We provide the data used in this work here. Table \ref{tab:table2} and Table \ref{tab:table3} show a part of the Kepler singles and Kepler multiples analysed in this work respectively. And Table \ref{tab:table4} and Table \ref{tab:table5} show a part of the RV singles and RV multiples analysed in this work respectively.}


\begin{deluxetable*}{crrrcc}
\tablenum{2}
\tablecaption{{\jw Kepler single planet systems analysed in this work.} \label{tab:table2}}
\tablehead{\colhead{Kepler Name} &\colhead{period} &\colhead{$R/R_{*}$} &\colhead{transit duration} &\colhead{SNR}&\colhead{...} \\ 
\colhead{} &\colhead{(day)} &\colhead{} &\colhead{(hour)} &\colhead{} &\colhead{...} 
}
\startdata
K00049.01&$8.313773_{-0.000042}^{+0.000042}$&$0.02636_{-0.00081}^{+0.00124}$&$2.964_{-0.125}^{+0.125}$&19.6& \\ 
K00072.02&$45.294223_{-0.000056}^{+0.000056}$&$0.01981_{-0.00013}^{+0.00033}$&$6.830_{-0.034}^{+0.034}$&134.8& \\ 
K00084.01&$9.286964_{-0.000004}^{+0.000004}$&$0.02378_{-0.00011}^{+0.00035}$&$3.365_{-0.015}^{+0.015}$&251.9&  
\enddata
    \begin{tablenotes}
      \small
      \item Note: Planet parameters, stellar parameters (except [Fe/H]) and [Fe/H] are from Kepler DR 25 \citep{koidr25}, \citet{2020AJ....159..280B} and LAMOST DR8 respectively. {\jw Here, only a part of table is shown and the whole table is available online.} 
    \end{tablenotes}

\end{deluxetable*}

\begin{deluxetable*}{crrrcc}
\tablenum{3}
\tablecaption{{\jw Kepler multiple planet systems analysed in this work.}\label{tab:table3}}
\tablehead{\colhead{Kepler Name} &\colhead{period} &\colhead{$R/R_{*}$} &\colhead{transit duration} &\colhead{SNR}&\colhead{...} \\ 
\colhead{} &\colhead{(day)} &\colhead{} &\colhead{(hour)} &\colhead{} &\colhead{...} 
}
\startdata
K00041.01&$12.815904_{-0.000019}^{+0.000019}$&$0.01407_{-0.00049}^{+0.00159}$&$6.386_{-0.044}^{+0.044}$&106.7& \\ 
K00041.02&$6.887071_{-0.000023}^{+0.000023}$&$0.00808_{-0.00063}^{+0.00001}$&$4.468_{-0.076}^{+0.076}$&38.2& \\ 
K00041.03&$35.333193_{-0.000225}^{+0.000225}$&$0.00962_{-0.00054}^{+0.00198}$&$5.966_{-0.128}^{+0.128}$&28.8&  
\enddata

    \begin{tablenotes}
      \small
      \item Note: Planet parameters, stellar parameters (except [Fe/H]) and [Fe/H] are from Kepler DR 25 \citep{koidr25}, \citet{2020AJ....159..280B} and LAMOST DR8 respectively. Here, only a part of table is shown and the whole table is available online.
    \end{tablenotes}
\end{deluxetable*}

\begin{deluxetable*}{crrrrr}
\tablenum{4}
\label{tab:table4}
\tablecaption{{\jw RV single planet systems analysed in this work.}}
\tablehead{\colhead{Planet Name} &\colhead{Radius} &\colhead{Mass(or Mass*$\sin i$)} &\colhead{e} &\colhead{Metallicity}&\colhead{Period} \\ 
\colhead{} &\colhead{(R$_{\oplus}$)} &\colhead{(M$_{\oplus}$)} &\colhead{} &\colhead{} &\colhead{(day)} 
}
\startdata
HD 97658 b&$ 2.12_{- 0.06}^{+ 0.06}$&$ 8.30_{- 1.10}^{+ 1.10}$&$0.050_{-0.030}^{+0.040}$&$-0.230_{-0.030}^{+0.030}$&$9.4897116_{-0.0000008}^{+0.0000008}$ \\ 
HD 95338 b&$ 3.89_{- 0.20}^{+ 0.19}$&$42.44_{- 2.08}^{+ 2.22}$&$0.197_{-0.024}^{+0.029}$&$0.040_{-0.100}^{+0.100}$&$55.0870000_{-0.0200000}^{+0.0200000}$ 
\enddata
    \begin{tablenotes}
      \small
      \item Note: Planet and stellar  parameters are from \citet{ps}. Here, only a part of table is shown and the whole table is available online.
    \end{tablenotes}
\end{deluxetable*}

\begin{deluxetable*}{crrrrr}
\tablenum{5}
\tablecaption{{\jw RV multiple planet systems analysed in this work.}\label{tab:table5}}
\tablehead{\colhead{Planet Name} &\colhead{Radius} &\colhead{Mass(or Mass*$\sin i$)} &\colhead{e} &\colhead{Metallicity}&\colhead{Period} \\ 
\colhead{} &\colhead{(R$_{\oplus}$)} &\colhead{(M$_{\oplus}$)} &\colhead{} &\colhead{} &\colhead{(day)} 
}
\startdata
HD 219134 f&$ 1.31_{- 0.02}^{+ 0.02}$&$ 7.30_{- 0.40}^{+ 0.40}$&$0.148_{-0.047}^{+0.047}$&$0.110_{-0.040}^{+0.040}$&$22.7170000_{-0.0150000}^{+0.0150000}$ \\ 
HD 219134 c&$ 1.51_{- 0.05}^{+ 0.05}$&$ 4.36_{- 0.22}^{+ 0.22}$&$0.062_{-0.039}^{+0.039}$&$0.110_{-0.040}^{+0.040}$&$6.7645800_{-0.0003300}^{+0.0003300}$ 
\enddata
    \begin{tablenotes}
      \small
      \item Note: Planet and stellar  parameters are from \citet{ps}. Here, only a part of table is shown and the whole table is available online.
    \end{tablenotes}
\end{deluxetable*}

\bibliography{ref_for_PET1}
\end{CJK*}
\end{document}